\documentclass[preprint,journal]{vgtc}       

\ifpdf
  \pdfoutput=1\relax                   
  \usepackage{graphicx}                
  \DeclareGraphicsExtensions{.pdf,.png,.jpg,.jpeg} 
\else
  \usepackage{graphicx}                
  \DeclareGraphicsExtensions{.eps}     
\fi%

\graphicspath{{figures/}{pictures/}{images/}{./}} 

\usepackage{microtype}                 
\PassOptionsToPackage{warn}{textcomp}  
\usepackage{textcomp}                  
\usepackage{mathptmx}                  
\usepackage{times}                     
\usepackage{cite}                      
\usepackage{tabu}                      
\usepackage{booktabs}                  

\usepackage{dirtytalk}
\usepackage{verbatimbox}
\usepackage{tikz}
\usetikzlibrary{shapes}

\usepackage{adjustbox}

\usepackage{graphicx}
\usepackage{subfig}

\usepackage{xspace} 
\usepackage{amssymb}

    \newcommand*\circled[1]{%
       \begin{tikzpicture}[baseline=(C.base)]
         \node[draw,circle,inner sep=0.3pt](C) {#1};
       \end{tikzpicture}}
       
    \newcommand*\boxed[1]{%
       \begin{tikzpicture}[baseline=(C.base)]
         \node[draw,rectangle,inner sep=0.6pt](C) {#1};
       \end{tikzpicture}}
    
    \newcommand*\textboxed[1]{%
       \begin{tikzpicture}[baseline=(C.base)]
         \node[draw,rectangle,inner sep=0.6pt](C) {#1};
       \end{tikzpicture}
    }
    
    \newcommand*\textcircle[1]{%
       \begin{tikzpicture}[baseline=(C.base)]
         \node[draw,circle,inner sep=0.3pt](C) {#1};
       \end{tikzpicture}}
    
    \definecolor{awesome}{rgb}{1.0, 0.13, 0.32}

\definecolor{action}{RGB}{60,160,208}
\definecolor{property}{RGB}{255,233,0}
\definecolor{header}{RGB}{255,255,255}
\definecolor{marked}{RGB}{181,0,0}
\definecolor{vumo}{RGB}{42,131,186}
\definecolor{draco}{RGB}{176,204,80}
\definecolor{nie}{RGB}{197,50,42}
\definecolor{pearls}{RGB}{255,166,12}
        
\ieeedoi{10.1109/TVCG.2021.3114687}

\onlineid{1308}

\vgtccategory{Research}
\vgtcpapertype{theory/model}

\title{Knowledge Rocks:\\ Adding Knowledge Assistance to Visualization Systems}

\author{Anna-Pia Lohfink, Simon D. Duque Anton, Heike Leitte \textit{Member, IEEE}, and Christoph Garth \textit{Member, IEEE}}
\authorfooter{
\item
 All authors are with Technische Universität Kaiserslautern, Germany. E-mails: lohfink@cs.uni-kl.de, simon.duque\_anton@posteo.de, leitte@cs.uni-kl.de, garth@cs.uni-kl.de
}

\shortauthortitle{Lohfink \MakeLowercase{\textit{et al.}}: Knowledge Rocks: Adding knowledge assistance to visualization systems}

\abstract{
We present Knowledge Rocks, an implementation strategy and guideline for augmenting visualization systems to knowledge-assisted visualization systems, as defined by the KAVA model. Visualization systems become more and more sophisticated. Hence, it is increasingly important to support users with an integrated knowledge base in making constructive choices and drawing the right conclusions. We support the effective reactivation of visualization software resources by augmenting them with knowledge-assistance. To provide a general and yet supportive implementation strategy, we propose an implementation process that bases on an application-agnostic architecture. This architecture is derived from existing knowledge-assisted visualization systems and the KAVA model. Its centerpiece is an ontology that is able to automatically analyze and classify input data, linked to a database to store classified instances. We discuss design decisions and advantages of the KR framework and illustrate its broad area of application in diverse integration possibilities of this architecture into an existing visualization system. In addition, we provide a detailed case study by augmenting an it-security system with knowledge-assistance facilities.

} 

\keywords{Knowledge-Assisted Visualization, Ontology, IT-Security.}

\CCScatlist{ 
 \CCScat{K.6.1}{Management of Computing and Information Systems}%
{Project and People Management}{Life Cycle};
 \CCScat{K.7.m}{The Computing Profession}{Miscellaneous}{Ethics}
}

\teaser{
  \centering
  \includegraphics[width=\linewidth]{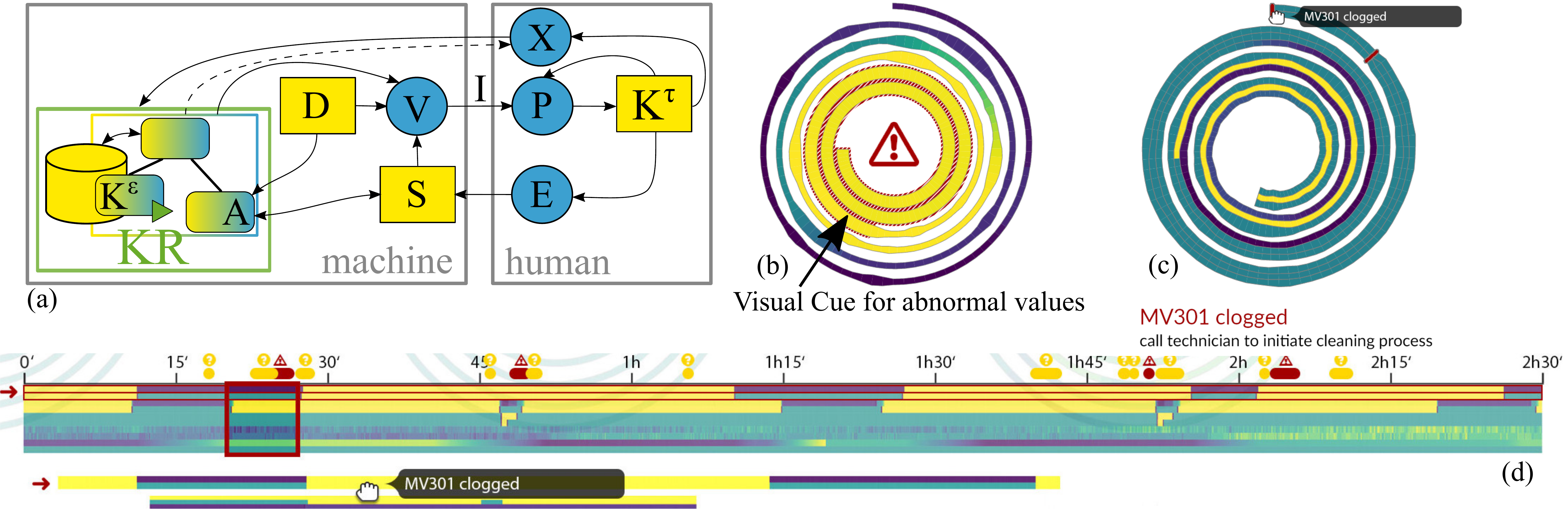}
  \caption{(a) The \textbf{Knowledge Rocks framework} provides the components that are required to make a visualization system knowledge-assisted according to the KAVA model. (b-d) The \textbf{knowledge-assisted Security in Process System} bases on the Knowledge Rocks framework. Visual cues for abnormal values (b) and instances from the knowledge base are added to the spiral plot of the analyzed data using a stream graph (c). The time slider is augmented to show related instances from the knowledge base (d).}
  \label{fig:teaser}
}

\vgtcinsertpkg

\begin{document}


\firstsection{Introduction}

\maketitle

Knowledge-assisted visualization systems use stored knowledge to support analysis tasks. This stored knowledge can range from different visualization strategies over best practices in a single visualization system to specific application knowledge. The provided support can be implemented as guidance regarding visualization settings or choices, automated analysis and pre-processing of the input, generation of examples and much more. 

The utility of knowledge-assisted visualization systems is well understood: already in the late 90's, Fujishiro et al. developed the GADGET system that supports users in choosing suitable visualization systems for their goals under given constraints (e.g. data properties) \cite{gadget}. Since then, the idea of knowledge assistance was further augmented and formalized. 
Chen et al. provided a solid theoretical basis \cite{chenTop10,chenInformationKnowledge,chenBook} and Federico et al. developed a conceptual model of knowledge-assisted visual analytics~--~the KAVA model \cite{explicitKnowledge}. Pike et al. declare knowledge-based interfaces as an important research challenge \cite{pike2009science}, and knowledge-assisted visualization was discussed in an issue of the \textit{IEEE Computer Graphics and Applications} journal \cite{journalKnowledge-assisted}. 

Also in the collaborative setting, knowledge-assisted visualization systems exhibit their strength: users are able to share their knowledge between each other, even if they are not using the system concurrently. Collaborative and cloud based visualization systems both have access to the knowledge of multiple users and are able to spread it among their users with a single, central knowledge base. 

Despite the usefulness and increasing necessity of knowledge assistance in visualization systems, to our knowledge there are few general guidelines or frameworks for its realization. Specific implementations pursue different approaches and are not generally applicable. Theoretical approaches, on the other hand, provide impulses concerning the development of new knowledge-assisted systems, but significant transfer must be performed on an ad hoc basis to translate theory into concrete implementations. Finally, fully implemented frameworks that add knowledge-assistance to visualization systems limit the possibilities of knowledge integration and interaction with the knowledge base.

With the Knowledge Rocks framework (KR framework), we aim to bridge theory and practice of knowledge-assisted visualization by simplifying the extension of existing visualization systems to become knowledge-assisted, allowing an effective reactivation of software resources in the visualization community. We determine components in the KAVA model that are essential for knowledge-assistance. By identifying these components in existing knowledge-assisted visualization systems, we gauge the range of applications that a generally applicable architecture needs to support. Based on this, we derive an application-agnostic architecture that provides all required components for knowledge assistance when integrated in a visualization system. 

As the centerpiece of the resulting KR framework, we define an \emph{acting ontology} as an ontology with callback functions. These functions are linked to ontology classes and allow automatic traversal of the ontology based on input data. To do so, a callback function decides to which neighbor of its linked class the traversal proceeds, based on the given input data. Hence, an acting ontology allows automatic analysis and classification of input data. It is used to retrieve, analyze and query classified instances that are stored in a database.

We discuss our design decisions and provide various integration possibilities of the KR framework into visualization systems. The resulting systems store different kinds of knowledge and provide different kinds of assistance, ranging from rule based reasoning to machine learning approaches. As a detailed case study, we augment the Security in Process System by Lohfink et al. to enable knowledge-assisted triage analysis \cite{securityinprocess}. We discuss in detail the concrete implementation and integration of the KR framework in the visualization system. 

After introducing technical terms and the KAVA model, we discuss related work in Section~\ref{relatedWork}. Subsequently, we make the following contributions: In Section~\ref{KR}, we present the Knowledge Rocks framework with the derivation of its requirements (\ref{derivation}), definition (\ref{architecture}), validation (\ref{validation}), and application to several examples (\ref{application}). Secondly, in a detailed case study, we integrate the KR framework in the Security in Process system and provide usage scenarios together with expert feedback on the new system in Section~\ref{tool}. We discuss the KR framework and its limitations in Section~\ref{discussion} and conclude with a summary of open questions and further opportunities in Section~\ref{conclusion}.

\section{Background and Related Work}\label{relatedWork}
The presented KR framework supports the extension of existing visualization systems to be knowledge-assisted. Its centerpiece is an ontology that is used for storing, processing and retrieval of knowledge. In the following, we provide some theoretical background and shortly discuss related work in relevant research areas.

Further related work pertaining to the application of the KR framework to the Security in Process System is discussed in  Section~\ref{implementationBackground}.

\subsection{Knowledge-Assisted Visualization}
Knowledge-assisted visualization aims to incorporate knowledge into the visualization process to support users \cite{chenBook}. 

Around the formalization of knowledge-assisted visualization systems \cite{chenInformationKnowledge} and the KAVA model\cite{explicitKnowledge}, a particular taxonomy evolved. We give a short overview of terms used in this paper: 

\textit{Tacit knowledge} contains a user's personal knowledge about data and insights gained during the perception of the visualization \cite{explicitKnowledge}. Knowledge that is available to assist the visualization forms the \textit{knowledge base}. To incorporate tacit knowledge into the knowledge base, it needs to be \say{written down} and become \textit{explicit knowledge}. This process is referred to as \textit{externalization}\cite{nonaka2007knowledge, WANG2009616}. Besides \textit{direct externalization} where a user is explicitly writing down their knowledge, automated externalization methods can be used that continuously extract knowledge in the background, for example from user interactions (\textit{interaction mining}). Different types of knowledge provide different support when they are leveraged in knowledge-assisted visualization: \textit{Operational knowledge} is about handling the visualization system and supports users in interacting with the visualization. \textit{Domain knowledge} contains knowledge about the analyzed data and helps users to interpret the content of the visualization. 

\subsection{The KAVA Model}\label{kavamodel}
\begin{figure}[t]
    \centering
    \includegraphics[width=0.55\columnwidth]{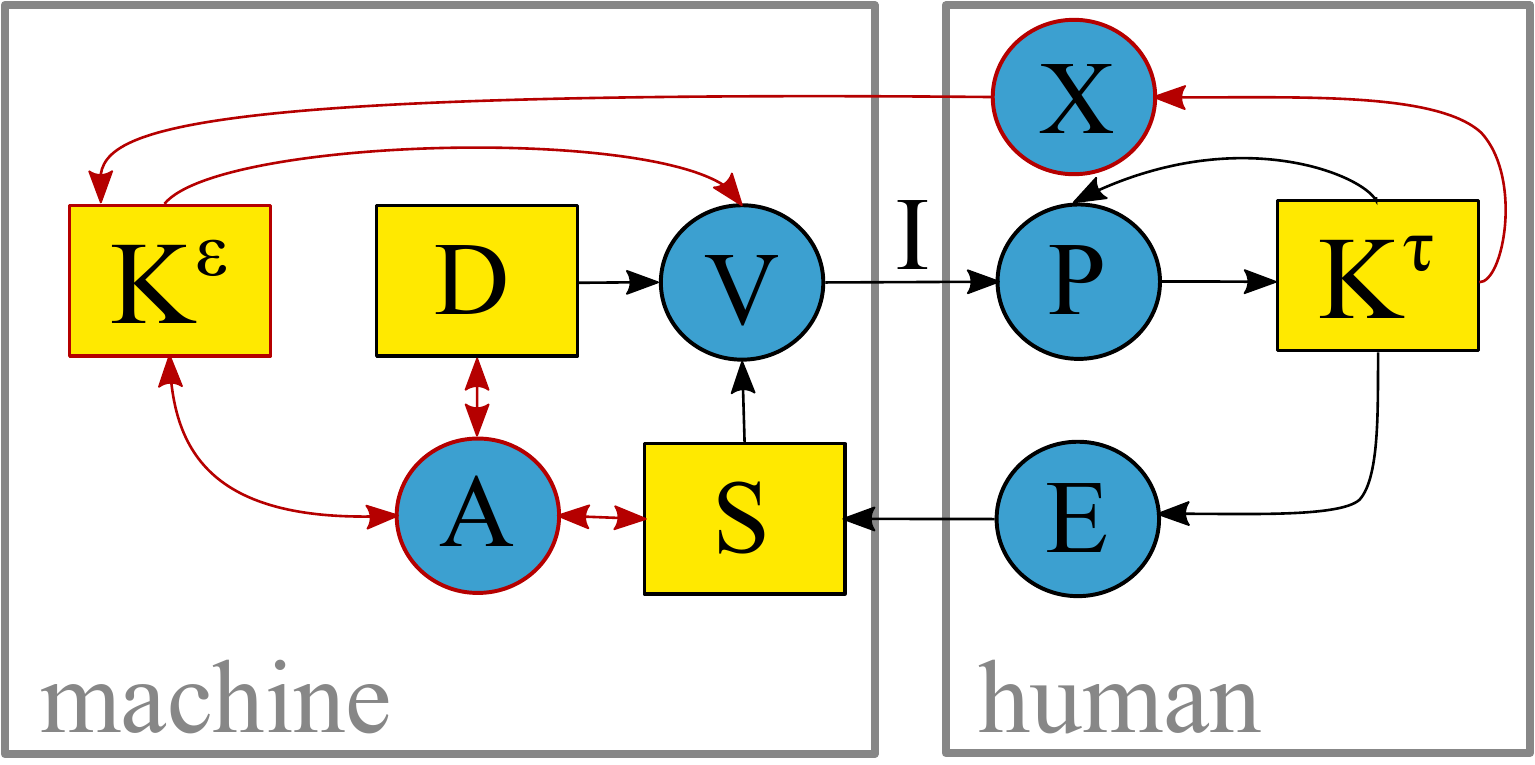}
    \caption[The KAVA model]{\label{fig:Model}\textbf{The KAVA model} of knowledge-assisted visual analytics by Federico et al. \cite{explicitKnowledge}. The processes: Analysis \textcircle{A}, visualization \textcircle{V}, externalization \textcircle{X}, perception/cognition \textcircle{P} and exploration \textcircle{E}; the containers: Explicit knowledge \textboxed{$K^{\varepsilon}$}, data \textboxed{D}, specification \textboxed{S}, tacit knowledge \textboxed{$K^{\tau}$}, and a non-persistent artifact: image I. Edges describe potential influences of data or knowledge stores and processes. Components and influences that distinguish a visualization system from a knowledge-assisted one are highlighted in red.}
\end{figure}

The \textit{KAVA model} by Federico et al. (Figure~\ref{fig:Model}) incorporates an explicit knowledge store and several knowledge-related processes in van Wijk's operational model of visualization~\cite{van2005value}. The basic knowledge assistance-related processes that are represented by their model are:

\emph{Knowledge visualization} \boxed{$K^{\varepsilon}$}$\rightarrow$\circled{V} Visualization \textcircle{V} of explicit knowledge \textboxed{$K^{\varepsilon}$} to present e.g. automatically extracted knowledge.

\emph{Simulation}
\boxed{$K^{\varepsilon}$}$\rightarrow$\circled{A}$\rightarrow$\boxed{D} External knowledge \textboxed{$K^{\varepsilon}$} is analyzed \textcircle{A} and applied to generate new data sets \textboxed{D} that provide users with supporting scenarios.

\emph{Automated/intelligent data analysis} (\boxed{D},~\boxed{S}$\,/\,$\boxed{$K^{\varepsilon}$} )$\rightarrow$\circled{A}$\rightarrow$\boxed{$K^{\varepsilon}$} The application of automated analysis \textcircle{A} to data \textboxed{D} to generate explicit knowledge \textboxed{$K^{\varepsilon}$}, given a certain specification \textboxed{S} or using explicit knowledge \textboxed{$K^{\varepsilon}$} respectively.

\emph{Direct externalization} \boxed{$K^{\tau}$}$\rightarrow$\circled{X}$\rightarrow$\boxed{$K^{\varepsilon}$} The system supports users in actively formulating tacit knowledge \textboxed{$K^{\tau}$} through an appropriate direct externalization interface \textcircle{X} to obtain explicit knowledge \textboxed{$K^{\varepsilon}$}.

\emph{Interaction mining}  \boxed{$K^{\tau}$}$\rightarrow$\circled{E}$\rightarrow$\boxed{S}$\rightarrow$\circled{A}$\rightarrow$\boxed{$K^{\varepsilon}$} Using their tacit knowledge \textboxed{$K^{\tau}$}, a user explores \textcircle{E} the data. The interaction with the system results in different specifications \textboxed{S} over time. These specifications are \say{mined}, that is they are automatically analyzed \textcircle{A} and contribute to the knowledge base \textboxed{$K^{\varepsilon}$}.

\emph{Guidance}
\boxed{$K^{\varepsilon}$}$\rightarrow$\circled{A}$\rightarrow$\boxed{S} Explicit knowledge \textboxed{$K^{\varepsilon}$} is analysed \textcircle{A} and used to guide the user's choice of settings \textboxed{S}. As described by Ceneda et al. \cite{guidance}, there are different degrees of guidance: Visual cues, providing alternative options and prescribing of specifications. An example for visual cues is the marking of abnormal high values in our exemplary implementation in Section~\ref{tool}.

The KAVA model aims to inspire innovative design approaches for knowledge-assisted visualization systems. Yet, creating such a system based on the model requires substantial transfer from the theoretical model to a practical implementation. With the KR framework, we lower this hurdle. To do so, we propose an application-agnostic architecture built around an ontology definition that, integrated in a visualization system, adds all functionality to the system to become knowledge-assisted in terms of the KAVA model.

\subsection{Knowledge-Assisted Visualization Systems}\label{knowledgeassistedsystems}

There is a large variety in knowledge-assisted visualization systems that have been developed so far and are described by the KAVA model. In the following, we present a representative selection of recent systems to illustrate the wide range of applications. We further use the presented systems as running examples to illustrate requirement analysis and application of the KR framework. 

1. \emph{VUMO: Towards an Ontology of Urban Mobility Events for Supporting Semi-Automatic Visualization Tools} \cite{VUMO}~--~Sobral et al. describe two ontologies that formalize the knowledge related to urban mobility events and visualizations respectively. They use this knowledge base to propose appropriate visualization techniques. Further, they provide a pipeline for a user-centered design process of visualization tools.

2. \emph{Knowledge-Assisted Comparative Assessment of Breast Cancer using Dynamic Contrast-Enhanced Magnetic Resonance Imaging} \cite{immerNie}~--~Nie et al. support physicians in exploration and classification of breast lesions. Lesions get scored based on their cluster structure and a knowledge base containing previously classified lesions using a fuzzy inference system. 

3. \emph{Formalizing Visualization Design Knowledge as Constraints: Actionable and Extensible Models in Draco} \cite{draco}~--~Moritz et al. implement a knowledge base containing visualization systems as well as hard constraints and weighted soft constraints concerning their application. Aiming on the acceleration of the transfer of research knowledge into practical tools, Draco guides users concerning visualization settings and choices.

4. \emph{KnowledgePearls: Provenance-Based Visualization Retrieval} \cite{stitz2018knowledgepearls}~--~Stitz et al. propose a system to build and visually access a provenance graph that stores visualization states and actions that occur during data analysis. They support direct access to the provenance graph based on queries (by selecting properties, and formulating statements in system or natural language) or based on created examples. 

All of these systems support data analysis: By proposing an adequate visualization (examples 1,3), by proposing visualization parameters (example 4) and by intelligent data analysis (example 2). A survey of 32 additional systems was given by Federico et al. \cite{explicitKnowledge}. 

\noindent\textbf{Ontologies} In computer science, \emph{ontologies} are a form of knowledge representation. Drawn as graphs, nodes of ontologies with degree one represent instances and nodes with higher degree represent classes or concepts. Edges represent the relationships among the classes with the most common relationship being \say{is a}. Ontologies are ideal to store knowledge about whole classes (\emph{class-based} knowledge) as opposed to \emph{case-based} knowledge about specific instances. They make stored knowledge accessible for both~--humans and computers--~because of their hierarchical structure. In addition, they have the ability to mutate as additional knowledge gets available. 

According to Carpendale et al. ontologies will be indispensable in developing infrastructures for knowledge-assisted visualization~\cite{BioOntologyVis}. This application is discussed for example by Miksch et al.~\cite{chenBook}; examples are the ontologies defined by Sobra et al. to support integration and visualization of data from intelligent transportation systems \cite{VUMO} and the three ontologies used by Gilson et al. to determine an appropriate visualization for web data (domain, visual representation and semantic bridging ontology)~\cite{gilson}. 

In our KR framework, the ontology is the starting point for an existing visualization system to become knowledge-assisted. It is used for structured storing, processing and retrieving of knowledge. 

We anticipate that knowledge-supported visual analysis will play a seminal role in future visualization systems and provide with the KR framework support for the effective reactivation of existing visualization systems. To demonstrate the high flexibility and generality of the proposed ontology-centered architecture, we motivate our design choices based on the running examples 1-4 in Section~\ref{derivation}, and sketch their possible implementation based on the KR framework in Section~\ref{application}.

\section{Knowledge Rocks Framework}\label{KR}
The KR framework provides a process to incorporate knowledge-assistance in a visualization system. It centers around an application-agnostic architecture that includes the existing visualization system as one of three components. Thus, the abstract process of knowledge incorporation boils down to the concrete implementation of components and establishing their interaction.
Within this implementation, all processes of a knowledge-assisted visualization system~--as described by the KAVA model--~can be reproduced. To become a component in an implementation that bases on the KR framework, an existing visualization system is only required to provide extensibility for the integration of the knowledge-assistance and interfaces for component interaction.

\subsection{Requirements}\label{derivation}
The KAVA model is a general model for knowledge-assisted visual analytics. Hence, it determines components that need to be added to a visualization system to make it knowledge-assisted. These components are:
explicit knowledge stored in a \emph{knowledge base} \textboxed{$K^\epsilon$} and \emph{automated analysis} \textcircle{A} using this knowledge base. An optional addition is support for \emph{direct externalization} \textcircle{X}. Required and optional components are colored in red in Figure~\ref{fig:Model}.

We examined several knowledge-assisted visualization systems with respect to their specific implementation of these components to obtain an application-agnostic architecture with a wide application range. In the following, we exemplary discuss the running examples 1-4:

1. VUMO's knowledge base consists of two ontologies~--a characterization of data and a characterization of visualizations--~and all integrated visualization systems and data sets. New data sets or queries are automatically analyzed by classifying them using the data ontology. The assigned perception factors are then used to link and suggest visualizations from the knowledge base. VUMO supports direct externalization by domain experts via the data ontology, and by visualization experts via the visualization ontology and the proposed visualization development pipeline. 

2. The knowledge base proposed by Nie et al. contains scored lesions with their linked clusters of time intensity curves, and linguistic rules that are created for each stored lesion. The main focus of this system lies on the automated analysis of dynamic contrast-enhanced magnetic resonance imaging data sets consisting of time intensity curve extraction, clustering, and scoring of lesions. Externalization of expert knowledge is supported by direct access to the knowledge base: scored data is presented together with stored results to support a possible correction of the scoring. Also, the human-readable linguistic rules for the scoring are available to users.

3. Draco's knowledge base consists of specifications of visualization systems and constraints. Its scope can be adapted depending on the application: Different visualization types or different specifications for a single visualization type can be stored. The input data~--data properties and an incomplete specification of the visualization--~is used to automatically suggest an optimal visualization from the knowledge base using cost functions with (optionally learned) weights. Hence, the automated analysis of the data is the optimal completion of the specification. Direct externalization by visualization experts is supported by providing a syntax to describe visualizations and constraints.

4. The knowledge base consists of a provenance graph together with stored properties. These properties are currently considered as independent; however, the authors state that additional knowledge concerning their dependencies could be integrated, which is part of their future work. The focus of KnowledgePearls lies on the direct access to the knowledge base, which is implemented with different query options. Thus, automated analysis of user interactions is kept simple: they are added to the provenance graph. The authors state that the size of the provenance graph is a limiting factor for externalization support. \\
\noindent
We extracted the following common patterns with respect to the required KAVA processes:

In all systems, the \emph{knowledge base} consists of two parts: A \say{passive} part with concrete instances (visualization systems in examples 1 and 3, lesions in 2, provenance in 4) and an \say{active} part consisting of rules or constraints acting on these instances (ontologies in example 1, linguistic rules in 2, constraints and cost functions in 3, stored properties and their dependencies in 4). This active part contains general relationships and reasoning used to classify input data. It corresponds to the \say{concepts} component in the structure defined by Rind et al. \cite{structuralFramework}.

In the first three examples, \emph{automated analysis} is implemented as the action of rules or constraints on the analyzed data and the knowledge base. 
In example 4 (KnowledgePearls), the stored properties are currently not subject to constraints. Thus, the automated analysis process does not rely on any rules or constraints. On the other hand, KnowledgePearls is the only system among the considered examples that focuses on direct access to the database. This access~--~via query by definition or example~--~follows the same pattern as the automated analysis in the other systems: given (and potentially stored) rules and constraints are applied to the knowledge base.

These observations~--the knowledge base consisting of a passive and an active part where the active part acts on the passive instances to provide automated analysis or direct access--~together with the structural definition of required and optional components in the KAVA model lead us to the following architecture definition.

\subsection{Architecture Definition}\label{architecture} 
\begin{figure}[t]
    \centering
    \includegraphics[width=0.75\linewidth]{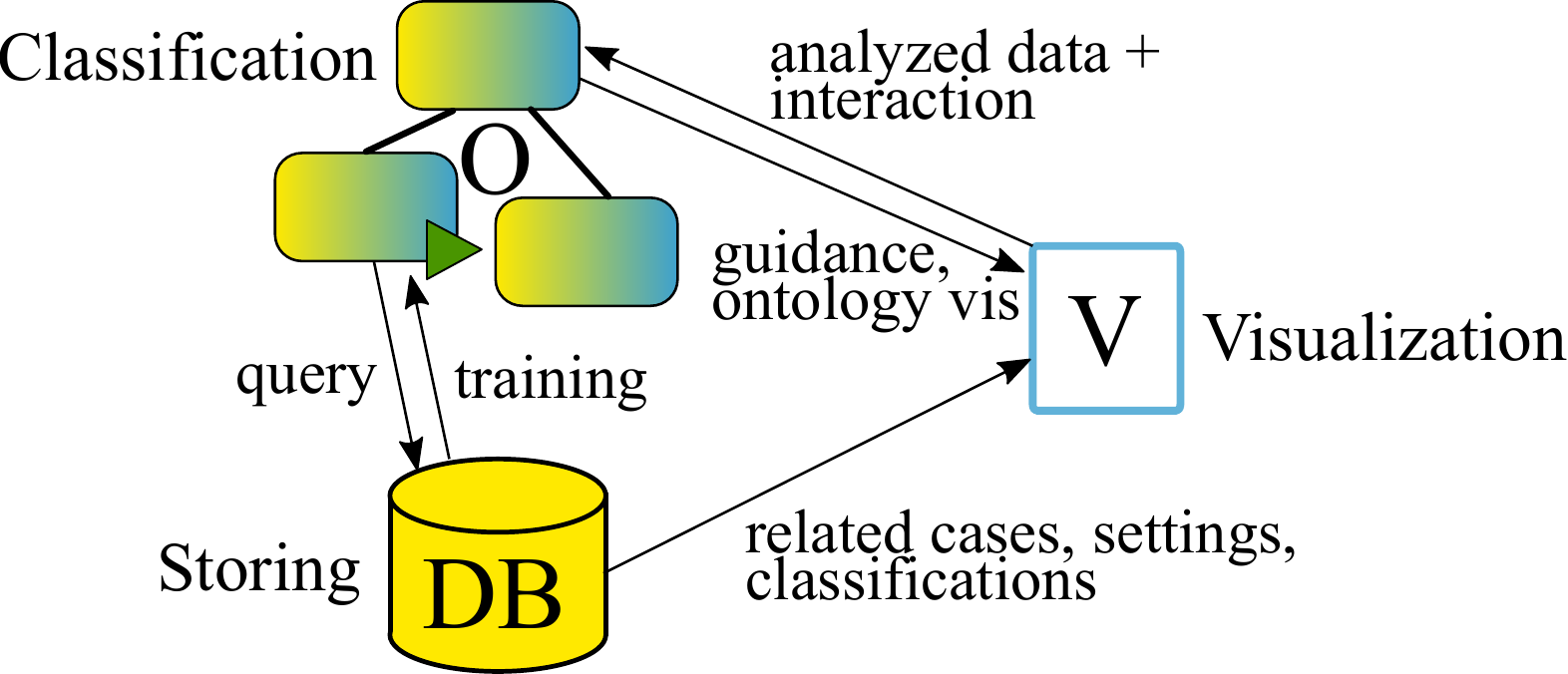}
    \caption{\label{fig:architecture}\textbf{Data flow} between the three parts of the KR framework: Two components are integrated in the visualization system (V): The acting ontology (O) that automatically classifies the analyzed data and the database (DB) that stores ontology instances and related structures.}
\end{figure}
The proposed architecture consists of two parts that are integrated into the existing \emph{visualization system}:
An \emph{acting ontology} and a supporting \emph{database} to store the ontology's instances and related data structures. See Figure~\ref{fig:architecture} for their interaction. 

This architecture implements the structure proposed by Rind et al. for domain knowledge in visual analytics \cite{structuralFramework}. Concepts are stored in the acting ontology and datasets in the database. The fulfillment of the desiderata they identified for structural models of domain knowledge are either inherent in our system (knowledge should be machine interpretable, pre-existing taxonomies can be used, implementation of the knowledge sources proposed in \cite{explicitKnowledge}, human readable knowledge, facilitating the exchange between different systems) or depend on the specific implementation (focussing on domain knowledge, compatibility with heterogeneous data, standardized form to include provenance information e.g. using triggers in the database) and the chosen language or tools for this implementation (software library support). In the following we present the different parts of our framework, give implementation and integration details, and discuss advantages and design choices. 


\noindent\textbf{Acting Ontology.}
A core aspect of the KR framework is the acting ontology (O in Figure~\ref{fig:architecture}). With this term we describe the combination of an ontology with a traversal strategy, implemented using callback functions. These functions are added to every class of the ontology. Reaching a class, the associated callback function identifies the next class on the traversing path based on the given input. Callback functions are formulated by domain experts and can include procedural reasoning, machine learning and other structures that decide which path of an ontology to follow and which thereby establish parent-child relations between the nodes. By traversing the acting ontology to an instance, the input data is classified as belonging to this instance. 

\begin{figure}[t!]
\centering
\adjustbox{valign=t}{
\scalebox{.8}{%
\begin{tikzpicture}[
    level 1/.style={sibling distance=1.5cm,level distance=1.0cm},
    level 2/.style={sibling distance=3.25cm, level distance=1.4cm},
    emph/.style={edge from parent/.style={marked,very thick,draw}},
    norm/.style={edge from parent/.style={black,thin,draw}}
    edge from parent path={(\tikzparentnode.south) -- (\tikzchildnode.north)},
    kant/.style={text width=2cm, text centered, sloped},
    every node/.style={text ragged, inner sep=0.5mm},
    punkt/.style={rectangle, rounded corners, draw=black!40!black!60, very
    thick }
    ]

\node[punkt,rectangle split, rectangle split parts=2, rectangle split part fill={header,action}] {
        \textbf{Class 1}
        \nodepart{second}
        $\texttt{callbackfunction()}$
    }
    child[emph] {
        node[punkt,rectangle split, rectangle split parts=2, rectangle split part fill={header,property}] {
            \textbf{Class 2}
            \nodepart{second}
            $\texttt{A}$
        }
        edge from parent
            node[below, pos=.6, edge from parent/.style = {draw=red}] {}
    }
    child {
        node[punkt,rectangle split, rectangle split parts=2, rectangle split part fill={header,property}] {
            \textbf{Class 3}
            \nodepart{second}
            $\texttt{B}$
        }
        edge from parent
            node[kant, below, pos=.6] {}
    }
    child {
        node[punkt,rectangle split, rectangle split parts=2, rectangle split part fill={header,property}] {
            \textbf{Class 4}
            \nodepart{second}
            $\texttt{C}$
        }
        edge from parent
            node[kant, below, pos=.6] {}
    };
\end{tikzpicture}}}
\hspace{0.5cm}
\adjustbox{valign=t}{
\scalebox{0.8}{
\begin{tikzpicture}[
    level 1/.style={sibling distance=1.5cm,level distance=1.0cm},
    level 2/.style={sibling distance=1.5cm, level distance=1.0cm},
    emph/.style={edge from parent/.style={marked,very thick,draw}},
    norm/.style={edge from parent/.style={black,thin,draw}},
    edge from parent path={(\tikzparentnode.south) -- (\tikzchildnode.north)},
    kant/.style={text width=2cm, text centered, sloped},
    every node/.style={text ragged, inner sep=0.5mm},
    punkt/.style={rectangle, rounded corners, draw=black!40!black!60, very
    thick }
    ]


\node[punkt,rectangle split, rectangle split parts=2, rectangle split part fill={header,action}] {
        \textbf{Class 1}
        \nodepart{second}
        $\texttt{callbackfunction()}$
    }
    child[emph] {
        node[punkt,rectangle split, rectangle split parts=2, rectangle split part fill={header,property}] {
            \textbf{Class 5}
            \nodepart{second}
            $\texttt{A},\texttt{B}$
        }
        child[emph] {
            node[punkt,rectangle split, rectangle split parts=2, rectangle split part fill={header,property}] {
                \textbf{Class 2}
                \nodepart{second}
                $\texttt{A}$
            }
            edge from parent
                node[kant, below, pos=.6] {}
        }
        child[norm] {
            node[punkt,rectangle split, rectangle split parts=2, rectangle split part fill={header,property}] {
                \textbf{Class 3}
                \nodepart{second}
                $\texttt{B}$
            }
            edge from parent
                node[kant, below, pos=.6] {}
        }
        edge from parent
            node[below, pos=.6, edge from parent/.style = {draw=red}] {}
    }
    child {
        node[punkt,rectangle split, rectangle split parts=2, rectangle split part fill={header,property}] {
            \textbf{Class 4}
            \nodepart{second}
            $\texttt{C}$
        }
        edge from parent
            node[kant, below, pos=.6] {}
    };
\end{tikzpicture}
}}
\caption{\label{fig:active_ontology}The \textbf{acting ontology}: The input data is classified as Class 2 if the callback function returns \texttt{A}. right: In case of a callback function that classifies more in-depth than the direct children, a list of possible results that is stored in a child's properties determines further traversal.}
\end{figure}
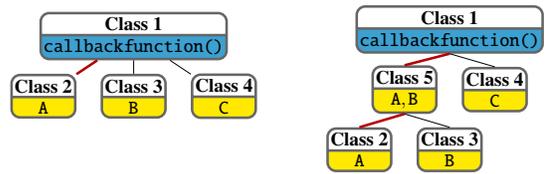

Concerning the concrete implementation of the acting ontology, we suggest the following: The ontology must to be interpretable by a computer; hence we implemented it in the web ontology language OWL \cite{owl_reference}. Paths in the ontology are coupled to results of the callback functions using ontology properties (cf.~Figure~\ref{fig:active_ontology}): A class of the acting ontology has a callback function as property. Every child of this class has a property listing one possible output of its parent's callback function, determining which child is the next class on the path. In some cases, a callback function is able to classify more in-depth than its direct children. In this case, a child class might not contain a callback function but a list of its children's properties. 

Examples for the structure of an acting ontology are given in Figure~\ref{fig:active_ontology}. Furthermore, our implementation of the acting ontology in the enhanced Security in Process System is given in Figure~\ref{fig:incident_onto}; its OWL code is given in the supplemental material.

We chose the ontology based approach for the following reasons:
Ontologies are a useful tool to capture and externalize knowledge. Their structure allows them to capture the active part of the stored knowledge. They are able to capture both, quantitative and nonquantitative knowledge, which is for example useful in Biology which is \say{notoriously nonquantitative} as pointed out by Carpendale et al. \cite{BioOntologyVis}. Also, they impose minimal boundaries to the designer~--who is able to choose classes, instances and their relations freely--~and still make the stored knowledge machine readable. 
The ability of ontologies to support knowledge externalization can be derived from their proximity to learning and brain storming techniques like mind maps \cite{mindmaps} and~--even closer--~concept maps \cite{conceptmap}. There is an active research field treating knowledge externalization techniques including ontologies: for example Ishikawa et al. propose an ontology based approach for knowledge externalization in companies \cite{ontologyforknowledge}, and Aranda-Corral et al. developed a tool that allows the collaborative development and improvement of an ontology \cite{ontoxicwiki}. All results from this area can be applied to design the ontology of the KR framework and develop it further.

Our choice to couple the ontology with callback functions to make it an acting ontology bases on several aspects:
To allow automated analysis and direct access to the knowledge base by applying the \say{active} part of the knowledge (stored in the ontology) to the stored instances, an acting part with executable rules and constraints is required. By incorporating this acting part directly in the ontology, we avoid a gap between concept and implementation. Knowledge by domain experts that is captured in the ontology is directly part of the knowledge-assisted visualization system. The most common relation between two classes in ontologies~--\say{is a}--~is in most cases easily translated in a callback function that tests if the given input \say{is an} instance of the different descendant classes. Designing more elaborate callback functions on the other hand requires the identification of logical flow between the defined classes and thus provides additional support for knowledge externalization. Finally, the acting ontology is machine \emph{and} human readable and thus presents a piece of self-documenting code that can further support knowledge externalization (for example the classification of instances that fail traversing the ontology) and the understanding of the visualization system. 

\noindent\textbf{Database.}
The database (DB in Figure~\ref{fig:architecture}) stores the \say{passive} instances and is tightly coupled to the acting ontology by the classification of instances as classes that are contained in the ontology. The nature of these instances is defined in the specific implementation, but we suggest to focus on domain knowledge as proposed by Rind et al. \cite{structuralFramework}. Whether the visualized data and the knowledge base reside in the same database is a design choice~--~depending on the stored knowledge, having both in the same database can avoid data amplification. Instances are stored automatically with all required additional information, after their classification by traversing the acting ontology. Hence, analyzed data that has been classified can be linked to data examples of the same class from the database. Stored data examples can contain operational data to support the visualization e.g. \say{All instances of the class \say{Contains Vortex} are visualized using streamlines}. They also can represent domain knowledge, e.g. typical patterns in the analyzed data. See Section~\ref{tool} for examples.

The database is required since our system is intended to potentially store numerous instances and storing them (and related data) in the acting ontology would make them difficult to access. In addition, a lot of background information and data points might be stored for an individual instance, blowing up the acting ontology. Databases are the intended tool for such applications, providing easy access to customizable data structures. The link to the acting ontology is given by the classification, that is the parent class in the ontology structure. 

\noindent\textbf{Visualization System.}
Finally, both entities~--acting ontology and database--~need to be coupled to the given visualization system (V in Figure~\ref{fig:architecture}). Besides automatic analysis that takes place in the background, there are two main types of interaction between users and the knowledge base in a knowledge-assisted visualization system: receiving guidance from the system and direct access to the knowledge base via knowledge visualization (as described in the KAVA model, Section~\ref{kavamodel}). In our framework, both types rely on the acting ontology.

\emph{Guidance} can help to narrow a user's knowledge gap by building upon various inputs like data, interaction history, stored domain knowledge and so on \cite{guidance}. The different guidance approaches~--Visual cues, providing alternative options, and prescribing of specifications--~can either depend on a whole class of instances (for example a visual cue: \say{All instances of the class \say{Abnormal high values} are visualized with a cue in the respective area}), or depend on individual instances (for example prescribing of specifications: \say{This specific example is visualized with a period of 3.5}). In the first case, they are triggered directly by the classification via the acting ontology: as soon as the provided data is ranked as a class with a property that triggers guidance, this trigger is passed to the visualization system and processed. In the latter case, guidance is offered if the instance that triggers the guidance is selected. An example for this is the optional prescription of stored specifications in Section~\ref{tool}.

Visual cues and alternative options require enhancements of the existing visualization system. While visual cues are incorporated directly in the visualization, alternative options are visualized stand alone, in a similar fashion as the analyzed data to allow a comparison. A straightforward possibility is to use the existing visualization not only for the analyzed data but also for related instances. Then, both visualizations can be combined using for example juxtaposition or superposition. Of course, more elaborate or application specific techniques can be applied and are in many cases preferable. A survey on comparative techniques in information visualization was given by Gleicher et al. \cite{infoVisComparison}. They also provide a design strategy for comparative visualization \cite{gleichercomparison}. In Scientific Visualization, comparative visualization of 3D ensembles was addressed by Demir et al. \cite{6875990}, Zhang et al. address tensor fields \cite{7192722} and Verma et al. present comparative flow visualization \cite{compflow}.

\emph{Direct access to the knowledge base} requires structures that allow filtering and querying of the stored data. Allowing to choose classes and paths in the acting ontology provides these structures. Using an interactive ontology visualization, users are able to browse different instances in the database by choosing classes until only currently helpful instances remain. The user's decision which path to take in the acting ontology can be supported by comments in the callback functions describing the reasoning behind the functions. These comments can be made accessible using a documentation generator. Furthermore, browsing the acting ontology can support users in manually classifying instances if the automated classification fails for some reason. There are multiple ontology browsers with different visualizations available: for example jambalaya \cite{jambalaya} and OntoViz \cite{ontoviz}. A recent survey on this topic was given by Dudas et al. \cite{ontovissurvey}.

An example for an acting ontology with callback functions, a database structure and the integration in a visualization system is given in Section~\ref{tool}.

\section{Implementation Steps for KAVA Processes with the KR Framework}\label{validation}
Via the step-by-step implementation of the KR framework in a concrete application, knowledge-assistance is added to the embedded visualization system.\\
1. \emph{Specify the knowledge} that builds the knowledge base, identify classes and instances and build the ontology and the database structure.\\
2. \emph{Add the active part} of the knowledge base: implement callback functions that allow a traversal of the ontology and hence automated classification and analysis.\\
3. \emph{Integrate} the knowledge base into the visualization system.
There are many possible interactions and processes that are typical for knowledge-assisted visualization systems and that can be implemented with the KR framework. In the following, we sketch the implementation of knowledge-assistance-related processes as described in the KAVA model (Section~\ref{kavamodel}):

\emph{Knowledge visualization} requires structures to browse the knowledge base. In the KR framework, the database can be queried by an interactive visualization of the acting ontology. Selecting specific classes, a user is able to restrict the obtained results and search for useful instances that are classified as one of the selected classes. Besides automatically offered guidance by the system, this visualized acting ontology allows users to get an overview of the available data in the knowledge base. The visualization of the result is provided by the visualization system, either stand alone or illustrative, in addition to the visualization of currently analyzed data.

\emph{Automated and intelligent data analysis} is performed by the acting ontology in form of automated classification. Processing and analysis steps are incorporated in the callback functions. 
To implement intelligent data analysis, the analysis and classification can be automatically improved by either changing the ontology or the callback functions, for example by training contained machine learning systems. 

\emph{Direct externalization} of tacit knowledge is supported during the implementation of the system by requiring the definition of the ontology and its callback functions. During operation of the knowledge-assisted visualization system, the ontology's structure, the documentation of its callback functions and the provided classifications further support knowledge externalization.

\emph{Interaction mining} and other automated knowledge generation processes are possible by automating the collection of specification data. After the collection was triggered, the collected data is automatically classified by the acting ontology and stored in the knowledge base.

\emph{Guidance} can be implemented as described in Section~\ref{architecture}, based on the classification of the currently analyzed data by the acting ontology. It can either be triggered by the class in the acting ontology directly, or by instances of the same class that are stored in the database.

\emph{Simulation} is a priori not a purpose of the KR framework. We focus on the analysis of existing data, not on creating new data. Nevertheless, an extension of the acting ontology-idea could be able to provide this, for example by linking functions that are able to generate data.

Thus, embedded in the KAVA model (Figure~\ref{fig:teaser}(a)), the KR framework keeps nearly the same connections as the knowledge-assisted specific components \textboxed{$K^\varepsilon$} and \textcircle{A} in the original model (Figure~\ref{fig:Model}). The edge \textcircle{A}$\rightarrow$\textboxed{D} is missing since the KR framework does a priori not support simulation. A symbolic edge from the acting ontology to the externalization process \textcircle{X} is added, emphasizing support for users in externalizing their tacit knowledge. This support is provided via visualization \textboxed{KR}$\rightarrow$\textcircle{V}, but also by providing a common taxonomy for all users. The proposed architecture provides the structure to build the knowledge-assisted visualization on. Possible interactions with the knowledge base in a concrete implementation strongly depend on the integration of the added components in the system and may differ substantially. Rind et al. give some typical approaches for the integration of knowledge in visualization systems \cite{structuralFramework} and discuss an example for knowledge assistance with direct knowledge access \cite{kavagait}. Ceneda et al. developed a framework to guide the development of knowledge-assisted visualization systems \cite{guideMe}.

\subsection{Application of the Framework}\label{application}
\definecolor{action}{RGB}{60,160,208}
\definecolor{property}{RGB}{255,233,0}
\definecolor{header}{RGB}{255,255,255}
\definecolor{marked}{RGB}{181,0,0}
\begin{figure}[t!]
\centering
\scalebox{.75}{%
\begin{tikzpicture}[
    level 1/.style={sibling distance=3cm,level distance=1.3cm},
    level 2/.style={sibling distance=3.0cm, level distance=1.4cm},
    level 3/.style={sibling distance=1.5cm, level distance=1.2cm},
    emph/.style={edge from parent/.style={marked,very thick,draw}},
    norm/.style={edge from parent/.style={black,thin,draw,inner sep=0pt,align=right,font=\scriptsize}},
    edge from parent path={(\tikzparentnode.south) -- (\tikzchildnode.north)},
    kant/.style={text width=2cm, text centered, sloped},
    every node/.style={text ragged, inner sep=0.5mm},
    punkt/.style={rectangle, rounded corners, draw=black!40!black!60, very
    thick }
    ]

\node[punkt,rectangle split, rectangle split parts=2, rectangle split part fill={header,action}] {
        \textbf{DCE-MRI Data Set}
        \nodepart{second}
        $\texttt{initialAnalysis()}$
    }
    child {
        node[punkt,rectangle split, rectangle split parts=2, rectangle split part fill={header,property}] {
            \textbf{No Lesion}
            \nodepart{second}
            $\texttt{"no lesion found"}$
        }
        edge from parent
            node[kant, above, pos=.75] {contains}
    }
    child {
        node[punkt,rectangle split, rectangle split parts=3, rectangle split part fill={header,property,action}] {
            \textbf{Lesion}
            \nodepart{second}
            $\texttt{"found lesion"}$
            \nodepart{third}
            $\texttt{evaluateRS()}$
        }
        child {
            node[punkt,rectangle split, rectangle split parts=3, rectangle split part fill={header,property,action}] {
                \textbf{Small Lesion}
                \nodepart{second}
                $\texttt{"small"}$
                \nodepart{third}
                $\texttt{FIS()}$
            }
            child {
                node[punkt,rectangle split, rectangle split parts=2, rectangle split part fill={header,property,action}] {
                    \textbf{Benign}
                    \nodepart{second}
                    $\texttt{0}\leq \texttt{score} \leq \texttt{2}$
                }
                edge from parent
                    node[kant, above, pos=.8] {is}
            }
            child {
                node[punkt,rectangle split, rectangle split parts=2, rectangle split part fill={header,property}] {
                    \textbf{...}
                    \nodepart{second}
                    $ $
                }
                edge from parent
                    node[kant, below, pos=.6] {}
            }
            child {
                node[punkt,rectangle split, rectangle split parts=2, rectangle split part fill={header,property,action}] {
                    \textbf{Malignant}
                    \nodepart{second}
                    $\texttt{6}\leq \texttt{score} \leq \texttt{8}$
                }
                edge from parent
                    node[kant, below, pos=.6] {}
            }
            edge from parent
                node[kant, below, pos=.6] {}
        }
        child {
            node[punkt,rectangle split, rectangle split parts=3, rectangle split part fill={header,property,action}] {
                \textbf{Medium Lesion}
                \nodepart{second}
                $\texttt{"medium"}$
                \nodepart{third}
                $\texttt{FIS()}$
            }
            child {
                node[punkt,rectangle split, rectangle split parts=2, rectangle split part fill={header,property}] {
                    \textbf{...}
                    \nodepart{second}
                    $ $
                }
                edge from parent
                    node[kant, below, pos=.6] {}
            }
            edge from parent
                node[kant, below, pos=.6] {}
        }
        child {
            node[punkt,rectangle split, rectangle split parts=3, rectangle split part fill={header,property,action}] {
                \textbf{Large Lesion}
                \nodepart{second}
                $\texttt{"large"}$
                \nodepart{third}
                $\texttt{FIS()}$
            }
            child {
                node[punkt,rectangle split, rectangle split parts=2, rectangle split part fill={header,property}] {
                    \textbf{...}
                    \nodepart{second}
                    $ $
                }
                edge from parent
                    node[kant, below, pos=.6] {}
            }
            edge from parent
                node[kant, below, pos=.6] {}
        }
        edge from parent
            node[kant, below, pos=.6] {}
    };
\end{tikzpicture}
}
    \caption{\label{fig:breastCancerOnto}Sketch of a possible \textbf{acting ontology for example 2}: The classes correspond to the linguistic variables. The analysis of DCE-MRI data sets is represented in this acting ontology: after initial analysis~--determining if there is a lesion detected or not--~the lesion size is determined based on the relative size (RS). The lesion score is then based on the Fuzzy Inference System \texttt{FIS}. The visualization of this ontology has no correspondence in their paper. Nonetheless, this would provide users a better understanding of classification options and boundaries in the system and a common terminology.}
\end{figure}
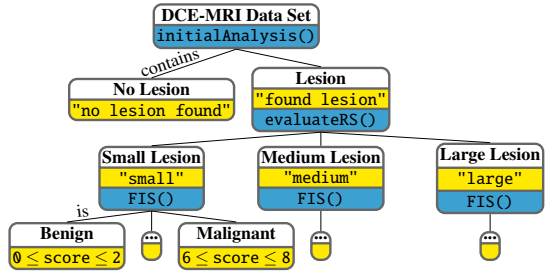
We briefly discuss possible implementations of the running examples 1-4 using the KR framework, reproducing the functionality of the examples based on an acting ontology with attached database. The various natures of these examples demonstrate the wide range of results that can be achieved with the framework. A detailed case study of an additional application is given in Section~\ref{tool}.

1. In VUMO, Sobral et al. define ontologies and discuss functions for automated classification, providing the callback functions in the acting ontology. The input data (a data set or the result of an analyzer) is classified by the data ontology resulting in a mobility event class, and spatial and temporal domains. With this classification set, the knowledge base can be queried to find an appropriate visualization based on the defined human perception factors. Integration of visualization systems is done using the visualization ontology.

2. The linguistic variables defined by Nie et al. provide the classes for an ontology definition (Figure~\ref{fig:breastCancerOnto}). After the initial analysis of the data using clusters from the database, the classification of a lesion's size is done via evaluation of the relative size (RS). To further classify the result in terms of lesion scores, the linguistic rules are evaluated. The storage structure of lesions and clusters is easily implemented in a database. Using the classification by the acting ontology, the database can be queried for similar lesions with the same clusters and classification. The visual integration of the system is as described in the paper. 

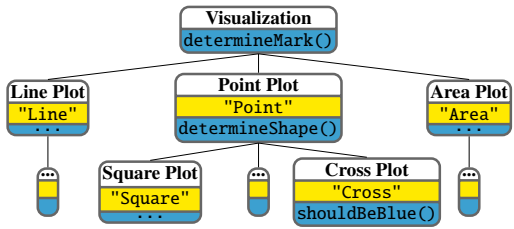
\begin{figure}[t]
\centering
\scalebox{.8}{%
\begin{tikzpicture}[
    level 1/.style={sibling distance=3.5cm,level distance=1.3cm},
    level 2/.style={sibling distance=1.8cm, level distance=1.4cm},
    emph/.style={edge from parent/.style={marked,very thick,draw}},
    norm/.style={edge from parent/.style={black,thin,draw}},
    edge from parent path={(\tikzparentnode.south) -- (\tikzchildnode.north)},
    kant/.style={text width=2cm, text centered, sloped},
    every node/.style={text ragged, inner sep=0.5mm},
    punkt/.style={rectangle, rounded corners, draw=black!40!black!60, very
    thick }
    ]


\node[punkt,rectangle split, rectangle split parts=2, rectangle split part fill={header,action}] {
        \textbf{Visualization}
        \nodepart{second}
        $\texttt{determineMark()}$
    }
    child {
        node[punkt,rectangle split, rectangle split parts=3, rectangle split part fill={header,property,action}] {
            \textbf{Line Plot}
            \nodepart{second}
            $\texttt{"Line"}$
            \nodepart{third}
            $\texttt{...}$
        }
        child {
            node[punkt,rectangle split, rectangle split parts=3, rectangle split part fill={header,property,action}] {
                \textbf{...}
                \nodepart{second}
                $ $
                \nodepart{third}
                $ $
            }
        }
    }
    child {
        node[punkt,rectangle split, rectangle split parts=3, rectangle split part fill={header,property,action}] {
            \textbf{Point Plot}
            \nodepart{second}
            $\texttt{"Point"}$
            \nodepart{third}
            $\texttt{determineShape()}$
        }
        child {
            node[punkt,rectangle split, rectangle split parts=3, rectangle split part fill={header,property,action}] {
                \textbf{Square Plot}
                \nodepart{second}
                $\texttt{"Square"}$
                \nodepart{third}
                $\texttt{...}$
            }
        }
        child {
            node[punkt,rectangle split, rectangle split parts=3, rectangle split part fill={header,property,action}] {
                \textbf{...}
                \nodepart{second}
                $ $
                \nodepart{third}
                $ $
            }
        }
        child {
            node[punkt,rectangle split, rectangle split parts=3, rectangle split part fill={header,property,action}] {
                \textbf{Cross Plot}
                \nodepart{second}
                $\texttt{"Cross"}$
                \nodepart{third}
                $\texttt{shouldBeBlue()}$
            }
        }
    }
    child {
        node[punkt,rectangle split, rectangle split parts=3, rectangle split part fill={header,property,action}] {
            \textbf{Area Plot}
            \nodepart{second}
            $\texttt{"Area"}$
            \nodepart{third}
            $\texttt{...}$
        }
        child {
            node[punkt,rectangle split, rectangle split parts=3, rectangle split part fill={header,property,action}] {
                \textbf{...}
                \nodepart{second}
                $ $
                \nodepart{third}
                $ $
            }
        }
    };
\end{tikzpicture}
}
    \caption{\label{fig:dracoOnto}Extract of a possible \textbf{acting ontology for example 3}, following the semantic tree of the Vega Lite syntax. A hard constraint is that only point plots can have a shape. \texttt{shouldBeBlue()} is an example for a soft constraints for point plots with cross shape. This callback allows different paths, potentially generating costs when the constraint is violated. A depth first search in the complete ontology corresponds to Draco's search. Although it is not implemented in Draco, showing this ontology to the user would provide an overview of possible options.}
\end{figure}

3. With the knowledge base consisting of visualization specifications in Vega Lite syntax, a possible ontology for Draco is similar to a semantic tree of this syntax. An example for this and for possible hard and soft constraints is given in Figure~\ref{fig:dracoOnto}. The callback functions deciding which soft constraints to follow or violate evaluate the cost functions. The ranked SVM discussed in \cite{draco} can be incorporated as such a callback function. Thus, the search implemented in Draco can be interpreted as depth first search in the tree structure of the ontology. 

4. For KnowledgePearls, the interaction provenance graph represents the knowledge base and is already stored in a database. While structuring the properties in an ontology is still under research, a simple tree structure representing one property per level can be used instead of an ontology. With this tree structure, the classification and querying of mined interactions is possible. Querying using natural language or SQL syntax can be implemented on top of the ontology. Fuzzy search is a priori not implementable using our system since searching by classification gives absolute results. The process can be mimiced by selecting multiple classes and rank the resulting instances based on the number of their ancestors that are selected. More details on this are given in the discussion of limitations (Section~\ref{discussion}). Query by example on the other hand is built in our framework: the user generated example is classified by the acting ontology and appropriate results from the database are presented.

The following tools can be used to implement the KR framework: The acting ontology can be created using Protégé \cite{protege} in Web Ontology Language (OWL). Software packages for loading OWL representations are readily available for a variety of environments. For example it can be loaded into Python using Owlready2 \cite{owlready} and into javaScript using owlreasoner \cite{owlreasoner}.
To keep the acting ontology easy to read and to support the implementation and debugging of the functions, only the references to callback functions are given in the ontology properties. Hence, an additional file with callback implementations needs to be provided. An example is given in the supplemental material.

\section{Case Study - Security in Process}\label{tool}
The Security in Process System (\textit{SiP System}) was developed by Lohfink et al. to visually support triage analysis in industrial process data \cite{securityinprocess}. Related fields of research are: anomaly detection in time series \cite{2020arXiv200400433B}, visualization in cyber security \cite{CSsurvey, vizsecproceedings} and time-series visualization \cite{Fang_2020, time_series_survey}. We employ the SiP System to provide a detailed study on how the KR framework is applied to extend it with knowledge assistance. After a short review of the original system, we follow the steps outlined in Section~\ref{validation}: First, we define the knowledge base in Section~\ref{SiPKnowledgeBase}. Then, we define structure and classes of the acting ontology and implement callback functions that allow an automatic traversal (Section~\ref{SiPActingOntology}). After the definition and implementation of the knowledge base's input and (automated) output in Sections~\ref{SiPsuggestingCases} and~\ref{SiPstoringCases}, we integrate the knowledge in the SiP System in Section~\ref{SiPVisualization}.  

\subsection{The Security in Process System}\label{implementationBackground}
\begin{figure}[t]
    \centering
    \includegraphics[width=\linewidth]{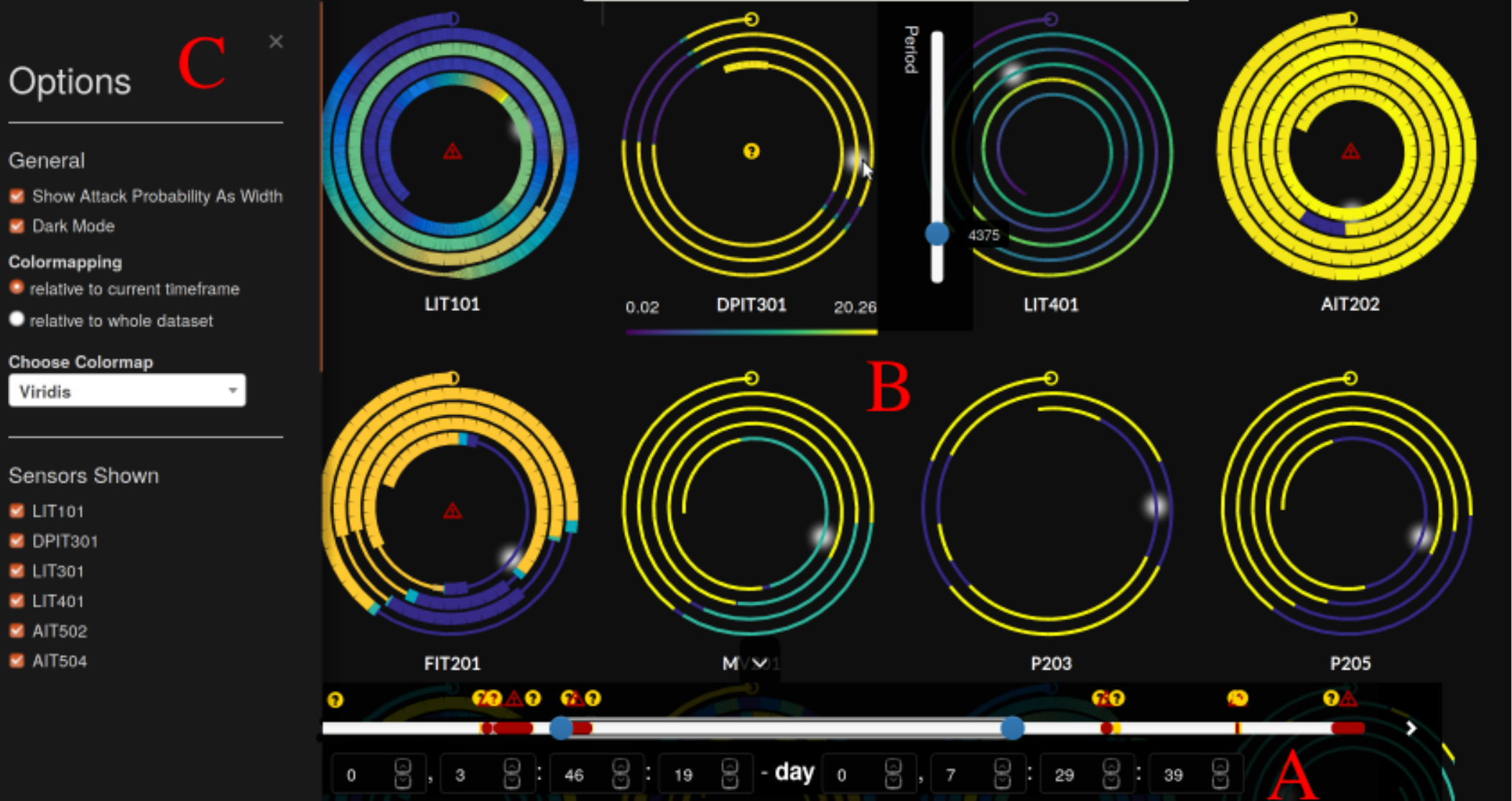}
    \caption{\label{fig:tool}\textbf{The Security in Process System} (SiP System) consisting of the time slider A, the spiral chart B and the options panel C. The time slider provides overview over the complete data set and indicates areas with warnings and alerts. In the spiral chart, readings and abnormality ratings of devices are shown by color and line thickness respectively.}
\end{figure}

In the SiP System, readings from sensors and actuators (\emph{devices}) involved in an industrial process are scanned for abnormal behavior (\emph{incidents}) using machine learning tools, resulting in an abnormality rating for every time step. This rating and the readings are shown to the analyst. To exploit the periodical behavior of industrial processes, spiral plots are used to encode the readings as color and the abnormality rating as line thickness (Figure~\ref{fig:tool} B).

Users can set the period of the individual spiral plots and choose the presented time frame in the time slider (Figure~\ref{fig:tool} A). This time slider represents the time frame of the complete data set, and indicates periods with potentially abnormal behavior using colored areas and symbols: warning \includegraphics[height=\fontcharht\font`\B]{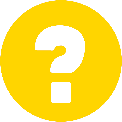} and alert \includegraphics[height=\fontcharht\font`\B]{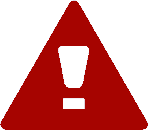}. On hovering a spiral plot with the cursor, a point highlights contemporaneous readings in the other plots, simplifying the comparison between plots with different periods.  

The used data set contains readings from PLCs monitoring a modern six-stage process of water treatment \cite{iTrust.2018,7469060}. Polluted water is pumped through the system, treated using different supplements in six stages and checked by different sensors until it is clean or re-enters the process.


\subsection{Knowledge Base}\label{SiPKnowledgeBase}
The SiP System is used by multiple analysts concurrently and/or asynchronous in a shift work schedule. Building a knowledge base can significantly improve the usefulness of the system, providing support for understanding and decisions, especially since the system addresses both experts and laymen.

As an anomaly detection system, the SiP System focuses on incidents that are defined as periods of the analyzed time series with a high abnormality rating. Shared knowledge can for example incorporate exemplary instances of proven attacks, instances of false positives in the anomaly detection, visualization settings for individual sensors or actuators, and visual cues for specific classes of incidents. Thus, the active knowledge stored in the acting ontology is based on incidents, and it is used to classify readings of devices with high abnormality rating. It is active in the sense that stored incidents from the database are described and classified by the knowledge in the acting ontology. 

The instances stored in the database represent the passive knowledge. They consist of readings of possibly multiple devices within a fixed time frame; they represent interesting values and patterns of different devices regarding one specific incident. Since different devices often correspond to different incident classes during a single incident, an instance in the database is identified by a set of classification labels. Additional information, like for example periods of the spiral plots and chosen color maps can be stored with a link to instances and individual devices. Based on this information, guidance is provided.
(The database schema is given in the supplementary material.)

\subsection{Acting Ontology}\label{SiPActingOntology}
\begin{figure}
    \centering
    \scalebox{.8}{%
\begin{tikzpicture}[
        level 1/.style={sibling distance=3cm,level distance=1.3cm},
        level 2/.style={sibling distance=3.6cm, level distance=1.5cm},
        level 3/.style={sibling distance=2.8cm, level distance=1.4cm},
        level 4/.style={sibling distance=2.0cm, level distance=1.4cm},
        level 5/.style={sibling distance=3.2cm}
        edge from parent/.style={very thick,draw=black!40!black!60,
            shorten >=0pt, shorten <=0pt},
        edge from parent path={(\tikzparentnode.south) -- (\tikzchildnode.north)},
        every node/.style={text ragged, inner sep=0.5mm},
        punkt/.style={rectangle, rounded corners, draw=black!40!black!60, very
        thick }
        ]

\node[punkt,rectangle split, rectangle split parts=2, rectangle split part fill={header,action}] {
        \textbf{Incident}
        \nodepart{second}
        $\texttt{callbackFalsePositiveCheck}()$
    }
    child {
        node[punkt,rectangle split, rectangle split parts=2, rectangle split part fill={header,property}] {
            \textbf{False Positive}
            \nodepart{second}
            $\texttt{True}$
        }
        edge from parent
            node[below, pos=.6] {}
    }
    child {
        node[punkt,rectangle split, rectangle split parts=3, rectangle split part fill={header,property,action}] {
            \textbf{Anomaly}
            \nodepart{second}
            $\texttt{False}$
            \nodepart{third}
            $\texttt{callbackAnomalyType}()$
        }
        child {
            node [punkt,rectangle split, rectangle split parts=3, rectangle split part fill={header,property,action}] {
                \textbf{Abnormal Values}
                \nodepart{second}
                $\texttt{"abnormal values"}$
                \nodepart{third}
                Type: \{\texttt{High},\texttt{Low}\}
            }
            edge from parent
                node[below, pos=.6] {}
        }
        child {
            node [punkt,rectangle split, rectangle split parts=3, rectangle split part fill={header,property,action}] {
                \textbf{Abnormal Occurrence}
                \nodepart{second}
                $\texttt{"abnormal occurrence"}$
                \nodepart{third}
                $\texttt{callbackPeriodicTest}()$
            }
            child{
                node[xshift=-2.5cm, punkt,rectangle split, rectangle split parts=2, rectangle split part fill = {header, property}] {
                    \textbf{Periodic}
                    \nodepart{second}
                    $\texttt{1\,,2}$
                }
                child{
                    node[punkt, text width =1.5cm, align=center, rectangle split, rectangle split parts=2, rectangle split part fill={header,property}] {
                        \textbf{Phase \\Shift}
                        \nodepart{second}
                        $\texttt{1}$
                    }
                    edge from parent
                        node[below, pos=.6] {}
                }
                child{
                    node[punkt, text width =1.5cm, align=center, rectangle split, rectangle split parts=2, rectangle split part fill={header,property}] {
                        \textbf{Frequency\\Change}
                        \nodepart{second}
                        $\texttt{2}$
                    }
                    edge from parent
                        node[below, pos=.6] {}
                }
                edge from parent
                    node[below, pos=.6] {}
            }
            child{
                node[xshift=-1cm, punkt,rectangle split, rectangle split parts=3, rectangle split part fill={header,property,action}] {
                    \textbf{Not Periodic}
                    \nodepart{second}
                    $\texttt{3}$
                    \nodepart{third}
                    $\texttt{callbackDisruptType}()$
                }
                child{
                    node[punkt,text width =1.5cm, align=center, rectangle split, rectangle split parts=2, rectangle split part fill={header,property}] {
                        \textbf{Pattern\\Disrupt}
                        \nodepart{second}
                        $\texttt{"pattern"}$
                    }
                    edge from parent
                        node[below, pos=.6] {}
                }
                child{
                    node[punkt,text width =1.5cm, align=center, rectangle split, rectangle split parts=2, rectangle split part fill={header,property}] {
                        \textbf{Period\\Disrupt}
                        \nodepart{second}
                        $\texttt{"period"}$
                    }
                    edge from parent
                        node[below, pos=.6] {}
                }
                edge from parent
                    node[below, pos=.6] {}
            }
            edge from parent
                node[below,  pos=.6] {}
        }
    };
\end{tikzpicture}
}
    \caption{The \textbf{incident based acting ontology} for the SiP System: Callback functions (blue) and properties (yellow) that are required for the traversal. The OWL code is given in the supplemental material.}
    \label{fig:incident_onto}
\end{figure}
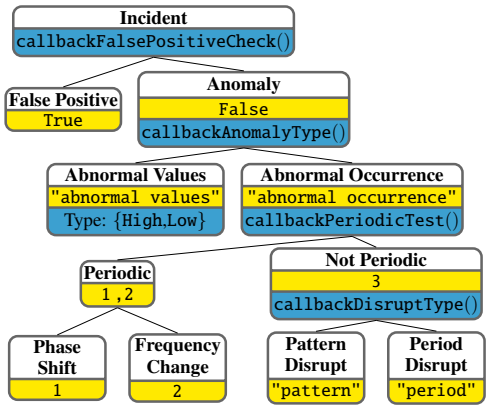

To classify incidents, we define the acting ontology as shown in Figure~\ref{fig:incident_onto} with the following classes:

\noindent\emph{Incident}: A series of readings that have been assigned a high abnormality rating; either an anomaly or false alert. 

\noindent\emph{False Positive and Anomaly}: As direct children of the incident class, these classes represent a false alert and the parent class for all anomaly types. An anomaly is either the occurrence of abnormal high or low values or the occurrence of values within the normal range at an unexpected time step ("Abnormal Occurrence"). 

\noindent\emph{Abnormal Values and Abnormal Occurrence}: Normal ranges for the individual devices can be determined using readings during incident-free operation. They are stored as sensor properties in the database. Anomalies are classified as abnormal values if they contain readings outside the normal range; if applicable, they are assigned the type \texttt{High} or \texttt{Low}. Abnormal occurrence of values means that they are within the normal range but do not follow the normal pattern of the considered time series (for example they disrupt an established period). Depending on the behavior of the system before and after the incident, we distinguish between periodic and not periodic abnormal occurrences. 

\noindent\emph{Periodic} and \emph{Not Periodic}: An abnormal occurrence of values is periodic if the considered time series is periodic before and after the incident with a possible transition at a constant value between two periods. The "Not Periodic" class describes all remaining abnormal occurrences of values.

\noindent\emph{Phase Shift and Frequency Change}: A periodic abnormal occurrence is a phase shift if the phase of the period is shifted and the frequency remains the same. If the frequency changes, it is a frequency change. 

\noindent\emph{Pattern and Period Disrupt}: A non-periodic abnormal occurrence is a period disrupt if an existing period is disrupted and does not re-occur. If there was no existing period, it is a pattern disrupt.

Note that to keep Figure~\ref{fig:incident_onto} comprehensible, additional properties are omitted. Especially for the anomaly instances (e.g. phase shift, frequency change, pattern disrupt and period disrupt), class-based remarks, actions to be taken and visualization settings such as visual cues can be added.

Our implementation of the callback functions includes pattern matching, machine learning components, and procedural reasoning to illustrate the variability of the KR framework. Details on their implementation and examples for the training
data and the generation approach are given in the supplemental material. 

\subsection{Suggesting Related Instances}\label{SiPsuggestingCases}
After classification of incidents in the analyzed data using the acting ontology, related instances from the database are automatically presented to users as decision support. The selection process works as follows: 
The classification of incidents results in the classification label set for all devices reporting an incident 
$\textit{Cl}_{\textit{cur}} = \{(c,cl)|c\,\, \textrm{device}, cl\,\, \textrm{classification}\}.$
In addition, the matrix profiles for every instance in the database and analyzed data have been calculated by \texttt{callbackFalsePositiveCheck}. For each device $c$ of each stored instance $\mathcal{I}$, the best matching position is stored together with the calculated distances $d_{\textit{min}}(c,\,\mathcal{I})$. Based on the set of classification labels and the best matching position for every case, the instances are ranked: $$\textit{rank}(\mathcal{I}) = \sum_{(c,\, cl) \in \textit{Cl}_{\textit{cur}} \bigcap \textit{Cl}_{\textit{db}}(\mathcal{I})} d_{\textit{min}}(c,\,\mathcal{I}).$$
With $\textit{Cl}_{\textit{db}}(\mathcal{I})$ being the set of classification labels of the currently ranked stored instance $\mathcal{I}$. The five instances with the highest ranking are then suggested in the enhanced time slider. 

\subsection{Storing Instances}\label{SiPstoringCases}
In case an analyst wants to store a new instance to the database, the storing mode can be entered by pressing a GUI button. The devices contained in the instance can be selected via mouse click. Their classification by the acting ontology is presented and users can change it using a drop down list containing all classes from the acting ontology. To get support for the classification, users can access the ontology visualization with the given class hierarchy and annotations from the callback functions.

In addition to the classification, annotations can be added for each device and for the whole instance. After entering a name for the instance, it is saved with a second click on the storing button. The stored properties are the selected devices and time frame with the corresponding data and anomaly ratings, annotations, and current visualization properties: period, selected color map and color map reference. The database schema and an illustration of the workflow is given in the supplemental material for more details.

\subsection{Visualization}\label{SiPVisualization}
As described in Section~\ref{architecture}, the visualization of the SiP System needs to be extended to incorporate knowledge assistance. We implemented the extended SiP System using D3.js \cite{d3} and python 3, framed by bottle \cite{bottle}. 

\begin{figure}[t!]
    \centering
    \includegraphics[width=\linewidth]{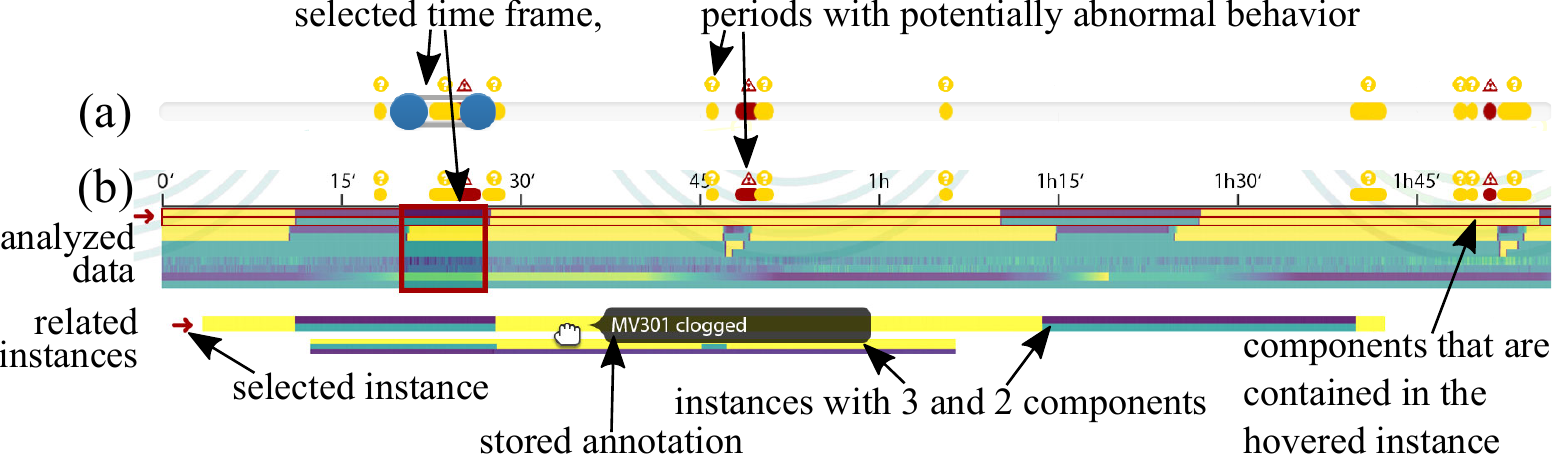}
    \caption{\textbf{Previous (a) and enhanced time slider (b)}: The line graph now gives an overview of all readings and related instances from the database. Optional clustering ensures similarity of adjacent readings.}
    \label{fig:tool_lineplot}
\end{figure}

\noindent\textbf{The Enhanced Time Slider.}
The time slider now includes information on the sensor readings and presents related instances from the database (Figure~\ref{fig:tool_lineplot}). To render this possible, we used a line graph following ideas by Kincaid et al. \cite{linegraph}. The analyzed data is shown on top, in a strand of device readings. Periods with potentially abnormal behavior are highlighted as in the previous version. Below the analyzed data related instances from the database are suggested with the same visual encoding and ordering. They can be selected, de-selected, and moved relative to the analyzed data with the mouse. The initial position of an instance is chosen according to the minimum distance as described in Section~\ref{SiPsuggestingCases}.
When an instance is hovered with the mouse, contained devices are highlighted in the time slider and the spiral chart; in addition, stored annotations for the hovered instance are shown as tool tip.

We optionally cluster the devices' readings to ensure that devices with similar patterns in their readings are adjacent in the time slider. The spiral plots in the spiral chart (Figure~\ref{fig:tool} B) are re-ordered accordingly. Forming flat clusters is achieved using the inconsistency method in SciPy \cite{scipy}. Similar to the previous version of the time slider, the selection frame can be re-positioned via drag and drop and its width can be changed with handles on the borders. 

\begin{figure}[t!]
    \centering
    \includegraphics[width=0.6\columnwidth]{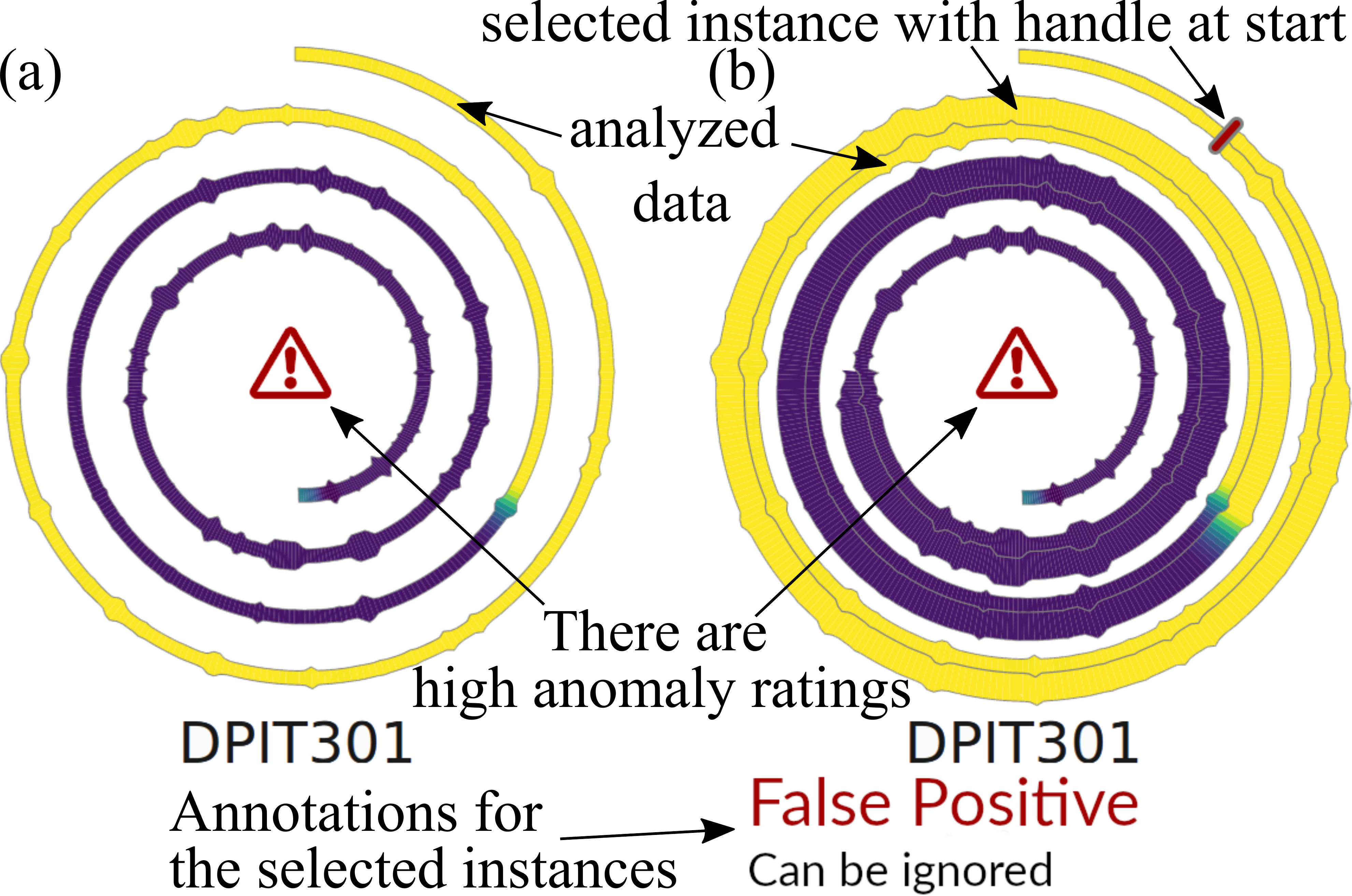}
    \caption{\textbf{Previous (a) and enhanced spiral plot (b)}: Readings can be compared and annotations provide decision support. \label{fig:tool_spiral}}
\end{figure}

\noindent\textbf{The Enhanced Spiral Plot.}
Selected instances are shown in the spiral chart: We extend the spiral plot by combining it with a stream graph (Figure~\ref{fig:teaser}(c) and~\ref{fig:tool_spiral}) similar to the approach by Jiang et al. \cite{jiang2016healthcare}. Using a stream graph that is centered at the spiral's center line, the thickness of the analyzed data and added instances accumulates. The analyzed data is always the innermost spiral. Handles at the beginning of a selected instance allow moving the instance with respect to the analyzed data; this is in addition to dragging the instance under the line graph. Annotations of the instance or its devices are shown on top of the spiral chart or below the according spiral plots respectively. Since not only the sequence of events but also their duration is important when searching for anomalies in industrial process data, selected instances are always shown with the same period as the analyzed data. Otherwise it would for example be possible to hide or create differences in the period.

Since spacing between spiral twists is an issue if multiple instances are selected or the anomaly rating is high, the maximum thickness for each strand is limited. If necessary, this limit is automatically reduced. In addition, it would be possible to change the selected period of the spiral to one of its integral multiples. Doing so, less twists are rendered but patterns are still recognizable. 

\noindent\textbf{Ontology Visualization.}
Direct access to the knowledge base is currently not required in the extended SiP System. Hence, the ontology visualization is only required to support the manual classification of instances. To do this, we implement a basic ontology visualization in a browsable tree layout similar to Figure~\ref{fig:incident_onto}. 

\subsection{Guidance}
Proposed incidents below the time slider can be chosen freely by the user, providing annotated examples to compare the current data with. Prescribing specifications can optionally be invoked by pressing the Alt-Key when selecting a related instance below the time slider. Then, linked visualization settings from the database are prescribed. A classification of an incident as \say{Abnormal Values} triggers highlighting of the period containing the abnormal values (Figure~\ref{fig:teaser}(b)). Additional visual cues supporting the identification of patterns could be implemented following the ideas by Ceneda et al. for spiral plots \cite{ceneda2018guided}.

The provided guidance is an optional addition to the SiP System. If none of the proposed incidents fits a situation where a user requires guidance (or there is no incident proposed), the browsable ontology and automated classification still provide support. In addition, users are urged to create a new incident in the knowledge base in such a case. 

\subsection{Data Size and Storage Access}\label{SiPperformance}
The considered data set consists of readings and anomaly ratings for 28 sensors at 467,919 time points. To create the enhanced time slider, every time step of all selected sensors needs to be loaded from the server. If static data is analyzed, this means that there is some seconds of loading time at system start. To keep this loading times small, the resolution of the transferred readings outside the selected time frame can be reduced. Changing the selected time frame, high resolution data for this frame is loaded dynamically. Streaming data, only added time steps need to be added to the right of the slider and cropped from the left, allowing fluent streaming. 

The knowledge base we created during the creation of the system consists of 26 incidents containing 23 different sensors over 615,844 time steps in total. Currently, at system start all incidents are loaded to ensure fast access times. With a growing knowledge base, more elaborate storage management is required, for example keeping incidents in memory based on their access frequency. During system use, stored incidents are accessed every time a new sensor is selected or the selected time frame is changed. Queries on the knowledge base during system use can be performed asynchronous in the background via ajax. By running a local server, access times are negligible. Remote access to the system might result in longer loading times. Classification of sensor readings via the acting ontology is done in milliseconds. Of course this runtime depends on the implementation of the callback functions. 

\subsection{Usage Scenarios and Expert Evaluation}

\begin{figure}[b!]
    \centering
    \includegraphics[width=0.8\columnwidth]{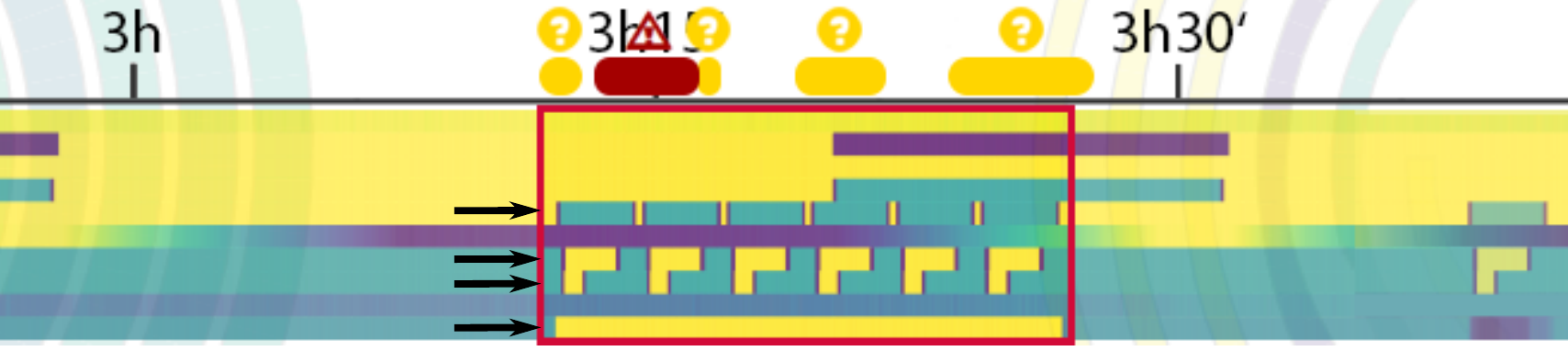}
    \caption{\label{fig:anomaly_in_linegraph}\textbf{Overview using the time slider}: Significant pattern changes can be spotted in the enhanced time slider.}
\end{figure}

We interviewed an it-security expert in the context of exemplary usage scenarios of the knowledge-assisted SiP System.

\noindent\textbf{Abnormal High Values.}
In the time slider in Figure~\ref{fig:anomaly_in_linegraph} an anomaly is indicated around 3h 10'. While the frequency change in the readings of several devices is quite obvious, the sensor at the bottom of the time slider faces an incident classified as \say{Abnormal Values} of type \say{High}. The linked visual cue highlights the period with abnormal high values in the spiral plot (Figure~\ref{fig:teaser}(b)). While the frequency changes are much more striking than the high values, the recorded attack actually originates on this device. The sensor was hacked, and the value was raised to provoke the recorded reaction in different actuators. 

The interviewed expert found it useful to be able to spot severe anomalies already in the time slider and independent of the results of the anomaly detection. According to him, the visual accentuation of different classified anomalies is a way to directly share knowledge with domain experts; visual cues are immediately visible and helpful. Especially values outside the normal range usually indicate severe incidents, be it a damage in the machine or an intrusion. Adding annotations to the stored incidents made sense to the expert because they enable laymen to react appropriately and support experts in troubleshooting. 

\noindent\textbf{False Positive Identification.}
In Figure~\ref{fig:tool_spiral}(a), the high abnormality rating for sensor DPIT301 triggers an alert, but there is no anomaly visible in the readings. In the evaluation of the SiP System in \cite{securityinprocess}, some users were reluctant to identify an incident as a false positive in a similar situation. The decision to disagree with the anomaly detection process is now supported by a related instance classified as "False Positive". Matching the analyzed data with the stored incidents in the database results in a very close match with this instance. Thus, it is suggested as related. Selecting it, the incident is added to the spiral plot at the calculated optimal position and the similarity is clearly visible (Figure~\ref{fig:tool_spiral}(b)). With this support, we expect it to be easier to identify false alerts by the anomaly detection system.

The intuitive incorporation of this very specific expert knowledge appeared fruitful to the interviewed expert. In his opinion, expanding the database with both, common and specific cases, to form a big knowledge base will be a huge support in triage analysis. 

\noindent\textbf{Creating and Using the Acting Ontology.}
The interviewed expert was enthusiastic to build a crucial part of the extended system as an ontology. In general, he finds it helpful to categorize and classify incidents, and comparatively easy to build the ontology. According to him, conducting research regarding the cause of an anomaly is facilitated by its classification. In his opinion, also the externalized expert knowledge on different anomaly types is helpful to get an overview, an idea of anomaly sources, and to manually classify anomalies. 

The expert highlighted positively that the ontology inherently introduces fundamental concepts that facilitate communication between experts and laymen using the system. Also, he saw great potential in the flexibility of the KR framework architecture, allowing for example one-class classifiers in the callback functions. 

\noindent\textbf{Further Development.}
As possible further development of the system, the expert recommended an automated prompt to add instances to the knowledge base where closely related instances are lacking, and to make the knowledge base (that is ontology and stored instances) accessible independently of the system for training purposes.

\section{Discussion and Limitations.}\label{discussion}
While the Knowledge Rocks framework provides an uncomplicated starting point for the extension of visualization systems to be knowledge-assisted, there are still some hurdles that need to be cleared for such an extension and some limits in the framework itself. 

\noindent\textbf{The ontology structure.} As the centerpiece of the KR framework, the ontology needs to be designed by experts and possibly with some effort. In some applications the structure of the knowledge to be stored is easy to find, but in other applications it might be a challenge to create an ontology structure. As a fallback, one can always use the tree structure representing one property per level instead of an ontology. Many callback functions need to be implemented in this case and browsing the tree to access the knowledge base is not as efficient as it could be with a carefully designed ontology. In addition, the design and implementation of callback functions potentially requires further expertise; however, we did not encounter substantially higher implementation complexity when using our framework compared to implementing the knowledge assistance from scratch. On the contrary, by directly incorporating results by domain experts, the implementation effort for developers can decrease.

\noindent\textbf{Query limitations.} Since the KR framework bases on the classification of analyzed data and instances, fuzzy search with different keywords is not possible. Nonetheless, searching for multiple properties at once is possible by allowing the selection of multiple classes in the ontology. Ranking of the results based on the number of selected parent classes can provide similar results to a fuzzy search. 

Since the classification is performed based on the ontology structure, it might be necessary to split especially real valued properties based on different value ranges. This prohibits searching for an exact value in the knowledge base. A solution for this is a second search in the search results of the ontology-query, searching for specific property values in the returned instances. 

\noindent\textbf{Required extension of the visualization.} To be as general as possible, the KR framework gives a basic structure and leaves the integration in the visualization to the users. For example the visualization needs to be extended to incorporate selected instances and to show instances from the knowledge base. Even the most simple approach using juxtaposition requires some implementation effort, especially if linking of the different visualizations is implemented. However, this effort is required also without using the KR framework.

\noindent\textbf{Misleading knowledge.} Misleading knowledge is one of the biggest issues in knowledge-assisted systems; if wrong information is added to the knowledge base, assistance can lead to wrong assumptions. While preventing this is only possible on user-side, one can still try to make the system more robust against such cases: in the knowledge-assisted SiP System, instances are proposed based on classification and pattern matching. If the classification of an incident in the knowledge base is wrong, it is likely not to show up in the top five. In addition, multiple instances are proposed. If annotations of one of them differs from all others, the user is likely to recognize the misleading information. In case of the ontology, different evaluation methods are available \cite{ontologyevaluation}.

\section{Conclusion and Future Work}\label{conclusion}
Knowledge-assisted visualization systems are playing an important role due to increasing complexity of both, data and visualization systems. With the KR framework, we support the extension of existing visualization systems to incorporate knowledge assistance. 

Based on the KAVA model, we determined components that are required to make a visualization system knowledge-assisted; then, we isolated these components in several knowledge-assisted systems and derived an application-agnostic architecture to provide them. After the validation of the KR framework using the KAVA model and giving implementation pointers, we applied it to several knowledge-assisted systems to demonstrate the wide range of possible results. Furthermore, we described in detail the integration of the KR framework in the SiP System and presented its advantages together with an expert interview. 

As future work, we will investigate more applications of our framework and possibilities to support the creation of an acting ontology on the structural and the implementation side. Also, we will explore the limitations of the KR framework with respect to the size and structure of the acting ontology, and possibilities to allow editing of the acting ontology by users, guided by ontology validation and possibly frameworks that support the creation of callback functions. 

Depending on the implementation of the callback functions, the classification via the acting ontology bases on causality or correlation; this can even differ between classes. Support to distinguish these in the KR framework and in the SiP System are an interesting research topic for the future. For example, the classification as \say{Abnormal Values} is causal, while all classifications made by machine learning approaches are based on correlation. 
Also handling misleading information that got into the knowledge base is an interesting research topic for KAVA systems in general, the KR framework, and concrete implementations. 

A practical evaluation of the SiP System and further development is planned; possibilities are direct access to the knowledge base for information and educational purposes, and the possibility to add unknown incidents with a \say{request} for annotations, that are then classified and annotated by experts.

\acknowledgments{
Funded by the Deutsche Forschungsgemeinschaft (DFG, German Research Foundation)~–~252408385~–~IRTG 2057}

\bibliographystyle{abbrv-doi}

\bibliography{template}

\begin{thebibliography}{10}

\bibitem{owlreasoner}
Javascript semantic web toolkit project page.
\newblock \url{https://code.google.com/archive/p/owlreasoner/}.
\newblock Accessed: 2020-03-29.

\bibitem{ontoxicwiki}
G.~A. Aranda-Corral, J.~Borrego-D{\'i}az, and A.~Jim{\'e}nez-Mavillard.
\newblock Social ontology documentation for knowledge externalization.
\newblock In {\em Metadata and Semantic Research}, pp. 137--148. Springer
  Berlin Heidelberg, Berlin, Heidelberg, 2010. doi: {{%
10\hspace{.1pt}\discretionary{.}{%
}{.}\hspace{.4pt}1007\discretionary{/}{%
}{/}978\discretionary{%
}{-}{-}3\discretionary{%
}{-}{-}642\discretionary{%
}{-}{-}16552\discretionary{%
}{-}{-}8\_14}}


\bibitem{owl_reference}
S.~Bechhofer, F.~{van Harmelen}, J.~Hendler, I.~Horrocks, D.~L. McGuinnes,
  P.~F. Patel-Schneider, and L.~A. Stein.
\newblock Owl web ontology language reference.
\newblock \url{https://www.w3.org/TR/owl-ref/}.
\newblock Accessed: 2020-11-25.

\bibitem{CSsurvey}
D.~Bhamare, M.~Zolanvari, A.~Erbad, R.~Jain, K.~Khan, and N.~Meskin.
\newblock Cybersecurity for industrial control systems: A survey.
\newblock {\em Computers \& Security}, 89:101677, 2020. doi: {{%
10\hspace{.1pt}\discretionary{.}{%
}{.}\hspace{.4pt}1016\discretionary{/}{%
}{/}j\hspace{.1pt}\discretionary{.}{%
}{.}\hspace{.4pt}cose\hspace{.1pt}\discretionary{.}{%
}{.}\hspace{.4pt}2019\hspace{.1pt}\discretionary{.}{%
}{.}\hspace{.4pt}101677}}


\bibitem{d3}
M.~Bostock, V.~Ogievetsky, and J.~Heer.
\newblock D$^3$ data-driven documents.
\newblock {\em IEEE Transactions on Visualization and Computer Graphics},
  17(12):2301--2309, 2011. doi: {{%
10\hspace{.1pt}\discretionary{.}{%
}{.}\hspace{.4pt}1109\discretionary{/}{%
}{/}TVCG\hspace{.1pt}\discretionary{.}{%
}{.}\hspace{.4pt}2011\hspace{.1pt}\discretionary{.}{%
}{.}\hspace{.4pt}185}}


\bibitem{2020arXiv200400433B}
M.~{Braei} and S.~{Wagner}.
\newblock {Anomaly Detection in Univariate Time-series: A Survey on the
  State-of-the-Art}.
\newblock {\em arXiv e-prints}, p. arXiv:2004.00433, 04 2020.

\bibitem{mindmaps}
T.~Buzan and B.~Buzan.
\newblock {\em The mind map book}.
\newblock Pearson Education, 2006.

\bibitem{BioOntologyVis}
S.~{Carpendale}, M.~{Chen}, D.~{Evanko}, N.~{Gehlenborg}, C.~{Görg},
  L.~{Hunter}, F.~{Rowland}, M.~{Storey}, and H.~{Strobelt}.
\newblock Ontologies in biological data visualization.
\newblock {\em IEEE Computer Graphics and Applications}, 34(2):8--15, 2014.
  doi: {{%
10\hspace{.1pt}\discretionary{.}{%
}{.}\hspace{.4pt}1109\discretionary{/}{%
}{/}MCG\hspace{.1pt}\discretionary{.}{%
}{.}\hspace{.4pt}2014\hspace{.1pt}\discretionary{.}{%
}{.}\hspace{.4pt}33}}


\bibitem{guideMe}
D.~Ceneda, N.~Andrienko, G.~Andrienko, T.~Gschwandtner, S.~Miksch,
  N.~Piccolotto, T.~Schreck, M.~Streit, J.~Suschnigg, and C.~Tominski.
\newblock Guide me in analysis: A framework for guidance designers.
\newblock {\em Computer Graphics Forum}, 39(6):269--288, 2020. doi: {{%
10\hspace{.1pt}\discretionary{.}{%
}{.}\hspace{.4pt}1111\discretionary{/}{%
}{/}cgf\hspace{.1pt}\discretionary{.}{%
}{.}\hspace{.4pt}14017}}


\bibitem{guidance}
D.~{Ceneda}, T.~{Gschwandtner}, T.~{May}, S.~{Miksch}, H.~{Schulz},
  M.~{Streit}, and C.~{Tominski}.
\newblock Characterizing guidance in visual analytics.
\newblock {\em IEEE Transactions on Visualization and Computer Graphics},
  23(1):111--120, 2017. doi: {{%
10\hspace{.1pt}\discretionary{.}{%
}{.}\hspace{.4pt}1109\discretionary{/}{%
}{/}TVCG\hspace{.1pt}\discretionary{.}{%
}{.}\hspace{.4pt}2016\hspace{.1pt}\discretionary{.}{%
}{.}\hspace{.4pt}2598468}}


\bibitem{ceneda2018guided}
D.~Ceneda, T.~Gschwandtner, S.~Miksch, and C.~Tominski.
\newblock Guided visual exploration of cyclical patterns in time-series.
\newblock \url{https://publik.tuwien.ac.at/files/publik_272789.pdf}, 10 2018.
\newblock poster presentation: Visualization in Data Science (VDS at IEEE VIS
  2018).

\bibitem{chenTop10}
C.~{Chen}.
\newblock Top 10 unsolved information visualization problems.
\newblock {\em IEEE Computer Graphics and Applications}, 25(4):12--16, 2005.
  doi: {{%
10\hspace{.1pt}\discretionary{.}{%
}{.}\hspace{.4pt}1109\discretionary{/}{%
}{/}MCG\hspace{.1pt}\discretionary{.}{%
}{.}\hspace{.4pt}2005\hspace{.1pt}\discretionary{.}{%
}{.}\hspace{.4pt}91}}


\bibitem{chenInformationKnowledge}
M.~{Chen}, D.~{Ebert}, H.~{Hagen}, R.~S. {Laramee}, R.~{van Liere}, K.~L. {Ma},
  W.~{Ribarsky}, G.~{Scheuermann}, and D.~{Silver}.
\newblock Data, information, and knowledge in visualization.
\newblock {\em IEEE Computer Graphics and Applications}, 29(1):12--19, 2009.
  doi: {{%
10\hspace{.1pt}\discretionary{.}{%
}{.}\hspace{.4pt}1109\discretionary{/}{%
}{/}MCG\hspace{.1pt}\discretionary{.}{%
}{.}\hspace{.4pt}2009\hspace{.1pt}\discretionary{.}{%
}{.}\hspace{.4pt}6}}


\bibitem{journalKnowledge-assisted}
M.~Chen and H.~Hagen.
\newblock Guest editors' introduction: Knowledge-assisted visualization.
\newblock {\em IEEE Computer Graphics and Applications}, 30(1):15--16, 01 2010.
  doi: {{%
10\hspace{.1pt}\discretionary{.}{%
}{.}\hspace{.4pt}1109\discretionary{/}{%
}{/}MCG\hspace{.1pt}\discretionary{.}{%
}{.}\hspace{.4pt}2010\hspace{.1pt}\discretionary{.}{%
}{.}\hspace{.4pt}8}}


\bibitem{6875990}
I.~{Demir}, C.~{Dick}, and R.~{Westermann}.
\newblock Multi-charts for comparative 3d ensemble visualization.
\newblock {\em IEEE Transactions on Visualization and Computer Graphics},
  20(12):2694--2703, 2014. doi: {{%
10\hspace{.1pt}\discretionary{.}{%
}{.}\hspace{.4pt}1109\discretionary{/}{%
}{/}TVCG\hspace{.1pt}\discretionary{.}{%
}{.}\hspace{.4pt}2014\hspace{.1pt}\discretionary{.}{%
}{.}\hspace{.4pt}2346448}}


\bibitem{ontovissurvey}
M.~{Dudáš}, S.~Lohmann, V.~Svátek, and D.~Pavlov.
\newblock Ontology visualization methods and tools: a survey of the state of
  the art.
\newblock {\em The Knowledge Engineering Review}, 33, 07 2018. doi: {{%
10\hspace{.1pt}\discretionary{.}{%
}{.}\hspace{.4pt}1017\discretionary{/}{%
}{/}S0269888918000073}}


\bibitem{Fang_2020}
Y.~Fang, H.~Xu, and J.~Jiang.
\newblock A survey of time series data visualization research.
\newblock {\em {IOP} Conference Series: Materials Science and Engineering},
  782:022013, 04 2020. doi: {{%
10\hspace{.1pt}\discretionary{.}{%
}{.}\hspace{.4pt}1088\discretionary{/}{%
}{/}1757\discretionary{%
}{-}{-}899x\discretionary{/}{%
}{/}782\discretionary{/}{%
}{/}2\discretionary{/}{%
}{/}022013}}


\bibitem{explicitKnowledge}
P.~Federico, M.~Wagner, A.~Rind, A.~Amor-Amor{\'o}s, S.~Miksch, and W.~Aigner.
\newblock The role of explicit knowledge: A conceptual model of
  knowledge-assisted visual analytics.
\newblock In {\em 2017 IEEE Conference on Visual Analytics Science and
  Technology (VAST)}, pp. 92--103, 2017. doi: {{%
10\hspace{.1pt}\discretionary{.}{%
}{.}\hspace{.4pt}1109\discretionary{/}{%
}{/}VAST\hspace{.1pt}\discretionary{.}{%
}{.}\hspace{.4pt}2017\hspace{.1pt}\discretionary{.}{%
}{.}\hspace{.4pt}8585498}}


\bibitem{vizsecproceedings}
F.~Fischer.
\newblock Vizsec proceedings browser.
\newblock \url{https://vizsec.dbvis.de/}, 2008.
\newblock Accessed: 2020-12-02.

\bibitem{gadget}
I.~{Fujishiro}, Y.~{Takeshima}, Y.~{Ichikawa}, and K.~{Nakamura}.
\newblock Gadget: goal-oriented application design guidance for modular
  visualization environments.
\newblock In {\em Proceedings. Visualization '97 (Cat. No. 97CB36155)}, pp.
  245--252, 1997. doi: {{%
10\hspace{.1pt}\discretionary{.}{%
}{.}\hspace{.4pt}1109\discretionary{/}{%
}{/}VISUAL\hspace{.1pt}\discretionary{.}{%
}{.}\hspace{.4pt}1997\hspace{.1pt}\discretionary{.}{%
}{.}\hspace{.4pt}663889}}


\bibitem{gilson}
O.~Gilson, N.~Silva, P.~Grant, and M.~Chen.
\newblock From web data to visualization via ontology mapping.
\newblock {\em Computer Graphics Forum}, 27(3):959--966, 2008. doi: {{%
10\hspace{.1pt}\discretionary{.}{%
}{.}\hspace{.4pt}1111\discretionary{/}{%
}{/}j\hspace{.1pt}\discretionary{.}{%
}{.}\hspace{.4pt}1467\discretionary{%
}{-}{-}8659\hspace{.1pt}\discretionary{.}{%
}{.}\hspace{.4pt}2008\hspace{.1pt}\discretionary{.}{%
}{.}\hspace{.4pt}01230\hspace{.1pt}\discretionary{.}{%
}{.}\hspace{.4pt}x}}


\bibitem{gleichercomparison}
M.~Gleicher.
\newblock Considerations for visualizing comparison.
\newblock {\em IEEE Transactions on Visualization and Computer Graphics},
  24(1):413--423, 2017. doi: {{%
10\hspace{.1pt}\discretionary{.}{%
}{.}\hspace{.4pt}1109\discretionary{/}{%
}{/}TVCG\hspace{.1pt}\discretionary{.}{%
}{.}\hspace{.4pt}2017\hspace{.1pt}\discretionary{.}{%
}{.}\hspace{.4pt}2744199}}


\bibitem{infoVisComparison}
M.~Gleicher, D.~Albers, R.~Walker, I.~Jusufi, C.~D. Hansen, and J.~C. Roberts.
\newblock Visual comparison for information visualization.
\newblock {\em Information Visualization}, 10(4):289--309, 2011. doi: {{%
10\hspace{.1pt}\discretionary{.}{%
}{.}\hspace{.4pt}1177\discretionary{/}{%
}{/}1473871611416549}}


\bibitem{bottle}
M.~Hellkamp et~al.
\newblock Bottle: Python web framework.
\newblock \url{https://bottlepy.org/docs/dev/}, 2016.
\newblock software, version 0.12.

\bibitem{ontologyforknowledge}
T.~Ishikawa, K.~Kobayashi, M.~Okabe, and T.~Yamaguchi.
\newblock Support for externalization of intelligence skill using ontology and
  rule-based system.
\newblock In {\em Proceedings of the 9th Joint Conference on Knowledge-Based
  Software Engineering, JCKBSE 2010}, Proceedings of the 9th Joint Conference
  on Knowledge-Based Software Engineering, JCKBSE 2010, pp. 145--159, Dec.
  2010.
\newblock 9th Joint Conference on Knowledge-Based Software Engineering, JCKBSE
  2010 ; Conference date: 25-08-2010 Through 27-08-2010.

\bibitem{iTrust.2018}
{iTrust Centre for Research in Cyber Security}.
\newblock Secure water treatment {(SWaT)} testbed.
\newblock Technical Report 4.2, Singapore University of Technology and Design,
  October 2018.

\bibitem{jiang2016healthcare}
S.~Jiang, S.~Fang, S.~Bloomquist, J.~Keiper, M.~J. Palakal, Y.~Xia, and S.~J.
  Grannis.
\newblock Healthcare data visualization: Geospatial and temporal integration.
\newblock In {\em VISIGRAPP (2: IVAPP)}, pp. 212--219, 01 2016. doi: {{%
10\hspace{.1pt}\discretionary{.}{%
}{.}\hspace{.4pt}5220\discretionary{/}{%
}{/}0005714002120219}}


\bibitem{linegraph}
R.~Kincaid and H.~Lam.
\newblock Line graph explorer: Scalable display of line graphs using
  focus+context.
\newblock In {\em Proceedings of the Working Conference on Advanced Visual
  Interfaces}, AVI '06, p. 404–411. Association for Computing Machinery, New
  York, NY, USA, 2006. doi: {{%
10\hspace{.1pt}\discretionary{.}{%
}{.}\hspace{.4pt}1145\discretionary{/}{%
}{/}1133265\hspace{.1pt}\discretionary{.}{%
}{.}\hspace{.4pt}1133348}}


\bibitem{owlready}
J.-B. Lamy.
\newblock Owlready2 documentation.
\newblock \url{https://owlready2.readthedocs.io/en/latest/}.
\newblock Accessed: 2020-11-27.

\bibitem{securityinprocess}
A.-P. Lohfink, S.~D.~D. Anton, H.~D. Schotten, H.~Leitte, and C.~Garth.
\newblock Security in process: Visually supported triage analysis in industrial
  process data.
\newblock {\em IEEE Transactions on Visualization and Computer Graphics},
  26(4):1638--1649, 2020. doi: {{%
10\hspace{.1pt}\discretionary{.}{%
}{.}\hspace{.4pt}1109\discretionary{/}{%
}{/}TVCG\hspace{.1pt}\discretionary{.}{%
}{.}\hspace{.4pt}2020\hspace{.1pt}\discretionary{.}{%
}{.}\hspace{.4pt}2969007}}


\bibitem{7469060}
A.~P. {Mathur} and N.~O. {Tippenhauer}.
\newblock {SWaT}: a water treatment testbed for research and training on {ICS}
  security.
\newblock In {\em 2016 International Workshop on Cyber-physical Systems for
  Smart Water Networks (CySWater)}, pp. 31--36, April 2016. doi: {{%
10\hspace{.1pt}\discretionary{.}{%
}{.}\hspace{.4pt}1109\discretionary{/}{%
}{/}CySWater\hspace{.1pt}\discretionary{.}{%
}{.}\hspace{.4pt}2016\hspace{.1pt}\discretionary{.}{%
}{.}\hspace{.4pt}7469060}}


\bibitem{chenBook}
S.~Miksch, H.~Leitte, and M.~Chen.
\newblock {\em Knowledge-Assisted Visualization and Guidance}, pp. 61--85.
\newblock Springer International Publishing, Cham, 2020. doi: {{%
10\hspace{.1pt}\discretionary{.}{%
}{.}\hspace{.4pt}1007\discretionary{/}{%
}{/}978\discretionary{%
}{-}{-}3\discretionary{%
}{-}{-}030\discretionary{%
}{-}{-}34444\discretionary{%
}{-}{-}3\_4}}


\bibitem{draco}
D.~{Moritz}, C.~{Wang}, G.~L. {Nelson}, H.~{Lin}, A.~M. {Smith}, B.~{Howe}, and
  J.~{Heer}.
\newblock Formalizing visualization design knowledge as constraints: Actionable
  and extensible models in draco.
\newblock {\em IEEE Transactions on Visualization and Computer Graphics},
  25(1):438--448, 2019. doi: {{%
10\hspace{.1pt}\discretionary{.}{%
}{.}\hspace{.4pt}1109\discretionary{/}{%
}{/}TVCG\hspace{.1pt}\discretionary{.}{%
}{.}\hspace{.4pt}2018\hspace{.1pt}\discretionary{.}{%
}{.}\hspace{.4pt}2865240}}


\bibitem{immerNie}
K.~Nie, P.~Baltzer, B.~Preim, and G.~Mistelbauer.
\newblock Knowledge-assisted comparative assessment of breast cancer using
  dynamic contrast-enhanced magnetic resonance imaging.
\newblock {\em Computer Graphics Forum}, 39(3):13--23, 2020. doi: {{%
10\hspace{.1pt}\discretionary{.}{%
}{.}\hspace{.4pt}1111\discretionary{/}{%
}{/}cgf\hspace{.1pt}\discretionary{.}{%
}{.}\hspace{.4pt}13959}}


\bibitem{nonaka2007knowledge}
I.~Nonaka and H.~Takeuchi.
\newblock The knowledge-creating company.
\newblock {\em Harvard business review}, 85(7/8):162, 2007.

\bibitem{conceptmap}
J.~D. Novak, D.~B. Gowin, and G.~D. Bob.
\newblock {\em Learning how to learn}.
\newblock cambridge University press, 1984.

\bibitem{protege}
N.~F. Noy, M.~Crub{\'e}zy, R.~W. Fergerson, H.~Knublauch, S.~W. Tu,
  J.~Vendetti, and M.~A. Musen.
\newblock Prot{\'e}g{\'e}-2000: An open-source ontology-development and
  knowledge-acquisition environment.
\newblock In {\em AMIA... Annual Symposium proceedings. AMIA Symposium}, p.
  953, 02 2003.

\bibitem{pike2009science}
W.~A. Pike, J.~Stasko, R.~Chang, and T.~O'connell.
\newblock The science of interaction.
\newblock {\em Information Visualization}, 8(4):263--274, 12 2009. doi: {{%
10\hspace{.1pt}\discretionary{.}{%
}{.}\hspace{.4pt}1057\discretionary{/}{%
}{/}ivs\hspace{.1pt}\discretionary{.}{%
}{.}\hspace{.4pt}2009\hspace{.1pt}\discretionary{.}{%
}{.}\hspace{.4pt}22}}


\bibitem{ontologyevaluation}
J.~Raad and C.~Cruz.
\newblock A survey on ontology evaluation methods.
\newblock In A.~L.~N. Fred, J.~L.~G. Dietz, D.~Aveiro, K.~Liu, and J.~Filipe,
  eds., {\em {KEOD} 2015 - Proceedings of the International Conference on
  Knowledge Engineering and Ontology Development, part of the 7th International
  Joint Conference on Knowledge Discovery, Knowledge Engineering and Knowledge
  Management {(IC3K} 2015), Volume 2, Lisbon, Portugal, November 12-14, 2015},
  pp. 179--186. SciTePress, 2015. doi: {{%
10\hspace{.1pt}\discretionary{.}{%
}{.}\hspace{.4pt}5220\discretionary{/}{%
}{/}0005591001790186}}


\bibitem{structuralFramework}
A.~Rind, M.~Wagner, and W.~Aigner.
\newblock Towards a structural framework for explicit domain knowledge in
  visual analytics.
\newblock In {\em 2019 IEEE Workshop on Visual Analytics in Healthcare (VAHC)},
  pp. 33--40, 2019. doi: {{%
10\hspace{.1pt}\discretionary{.}{%
}{.}\hspace{.4pt}1109\discretionary{/}{%
}{/}VAHC47919\hspace{.1pt}\discretionary{.}{%
}{.}\hspace{.4pt}2019\hspace{.1pt}\discretionary{.}{%
}{.}\hspace{.4pt}8945032}}


\bibitem{time_series_survey}
S.~F. {Silva} and T.~{Catarci}.
\newblock Visualization of linear time-oriented data: A survey.
\newblock In {\em Proceedings of the First International Conference on Web
  Information Systems Engineering}, vol.~1, pp. 310--319 vol.1, 2000. doi: {{%
10\hspace{.1pt}\discretionary{.}{%
}{.}\hspace{.4pt}1109\discretionary{/}{%
}{/}WISE\hspace{.1pt}\discretionary{.}{%
}{.}\hspace{.4pt}2000\hspace{.1pt}\discretionary{.}{%
}{.}\hspace{.4pt}882407}}


\bibitem{ontoviz}
G.~Singh, T.~Prabhakar, J.~Chatterjee, V.~Patil, S.~Ninomiya, et~al.
\newblock {OntoViz}: Visualizing ontologies and thesauri using layout
  algorithms.
\newblock In {\em The Fifth International Conference of the Asian Federation
  for Information Technology in Agriculture (AFITA 2006)}, 2006.

\bibitem{VUMO}
T.~Sobral, T.~Galv{\~a}o, and J.~Borges.
\newblock An ontology-based approach to knowledge-assisted integration and
  visualization of urban mobility data.
\newblock {\em Expert Systems with Applications}, 150:113260, 2020. doi: {{%
10\hspace{.1pt}\discretionary{.}{%
}{.}\hspace{.4pt}1016\discretionary{/}{%
}{/}j\hspace{.1pt}\discretionary{.}{%
}{.}\hspace{.4pt}eswa\hspace{.1pt}\discretionary{.}{%
}{.}\hspace{.4pt}2020\hspace{.1pt}\discretionary{.}{%
}{.}\hspace{.4pt}113260}}


\bibitem{stitz2018knowledgepearls}
H.~Stitz, S.~Gratzl, H.~Piringer, T.~Zichner, and M.~Streit.
\newblock Knowledgepearls: Provenance-based visualization retrieval.
\newblock {\em IEEE Transactions on Visualization and Computer Graphics (VAST
  '18)}, 25(1):120--130, 2018. doi: {{%
10\hspace{.1pt}\discretionary{.}{%
}{.}\hspace{.4pt}1109\discretionary{/}{%
}{/}TVCG\hspace{.1pt}\discretionary{.}{%
}{.}\hspace{.4pt}2018\hspace{.1pt}\discretionary{.}{%
}{.}\hspace{.4pt}2865024}}


\bibitem{jambalaya}
M.-A. Storey, M.~Musen, J.~Silva, C.~Best, N.~Ernst, R.~Fergerson, and N.~Noy.
\newblock Jambalaya: Interactive visualization to enhance ontology authoring
  and knowledge acquisition in prot{\'e}g{\'e}.
\newblock In {\em Workshop on Interactive Tools for Knowledge Capture},
  vol.~73, 12 2001.

\bibitem{scipy}
P.~Virtanen, R.~Gommers, T.~E. Oliphant, M.~Haberland, T.~Reddy, D.~Cournapeau,
  E.~Burovski, P.~Peterson, W.~Weckesser, J.~Bright, S.~J. {van der Walt},
  M.~Brett, J.~Wilson, K.~J. Millman, N.~Mayorov, A.~R.~J. Nelson, E.~Jones,
  R.~Kern, E.~Larson, C.~J. Carey, {\.I}.~Polat, Y.~Feng, E.~W. Moore,
  J.~{VanderPlas}, D.~Laxalde, J.~Perktold, R.~Cimrman, I.~Henriksen, E.~A.
  Quintero, C.~R. Harris, A.~M. Archibald, A.~H. Ribeiro, F.~Pedregosa, P.~{van
  Mulbregt}, and {SciPy 1.0 Contributors}.
\newblock {SciPy} 1.0: Fundamental algorithms for scientific computing in
  python.
\newblock {\em Nature Methods}, 17:261--272, 2020. doi: {{%
10\hspace{.1pt}\discretionary{.}{%
}{.}\hspace{.4pt}1038\discretionary{/}{%
}{/}s41592\discretionary{%
}{-}{-}019\discretionary{%
}{-}{-}0686\discretionary{%
}{-}{-}2}}


\bibitem{compflow}
{Vivek Verma} and A.~{Pang}.
\newblock Comparative flow visualization.
\newblock {\em IEEE Transactions on Visualization and Computer Graphics},
  10(6):609--624, 2004. doi: {{%
10\hspace{.1pt}\discretionary{.}{%
}{.}\hspace{.4pt}1109\discretionary{/}{%
}{/}TVCG\hspace{.1pt}\discretionary{.}{%
}{.}\hspace{.4pt}2004\hspace{.1pt}\discretionary{.}{%
}{.}\hspace{.4pt}39}}


\bibitem{kavagait}
M.~Wagner, D.~Slijepcevic, B.~Horsak, A.~Rind, M.~Zeppelzauer, and W.~Aigner.
\newblock Kavagait: Knowledge-assisted visual analytics for clinical gait
  analysis.
\newblock {\em {IEEE} Transactions on Visualization and Computer Graphics},
  25(3):1528--1542, 2019. doi: {{%
10\hspace{.1pt}\discretionary{.}{%
}{.}\hspace{.4pt}1109\discretionary{/}{%
}{/}TVCG\hspace{.1pt}\discretionary{.}{%
}{.}\hspace{.4pt}2017\hspace{.1pt}\discretionary{.}{%
}{.}\hspace{.4pt}2785271}}


\bibitem{WANG2009616}
X.~Wang, D.~H. Jeong, W.~Dou, S.-W. Lee, W.~Ribarsky, and R.~Chang.
\newblock Defining and applying knowledge conversion processes to a visual
  analytics system.
\newblock {\em Computers \& Graphics}, 33(5):616 -- 623, 2009. doi: {{%
10\hspace{.1pt}\discretionary{.}{%
}{.}\hspace{.4pt}1016\discretionary{/}{%
}{/}j\hspace{.1pt}\discretionary{.}{%
}{.}\hspace{.4pt}cag\hspace{.1pt}\discretionary{.}{%
}{.}\hspace{.4pt}2009\hspace{.1pt}\discretionary{.}{%
}{.}\hspace{.4pt}06\hspace{.1pt}\discretionary{.}{%
}{.}\hspace{.4pt}004}}


\bibitem{van2005value}
V.~Wijk.
\newblock The value of visualization.
\newblock In {\em VIS 05. IEEE Visualization, 2005.}, pp. 79--86, 2005. doi:
  {{%
10\hspace{.1pt}\discretionary{.}{%
}{.}\hspace{.4pt}1109\discretionary{/}{%
}{/}VISUAL\hspace{.1pt}\discretionary{.}{%
}{.}\hspace{.4pt}2005\hspace{.1pt}\discretionary{.}{%
}{.}\hspace{.4pt}1532781}}


\bibitem{7192722}
C.~{Zhang}, T.~{Schultz}, K.~{Lawonn}, E.~{Eisemann}, and A.~{Vilanova}.
\newblock Glyph-based comparative visualization for diffusion tensor fields.
\newblock {\em IEEE Transactions on Visualization and Computer Graphics},
  22(1):797--806, 2016. doi: {{%
10\hspace{.1pt}\discretionary{.}{%
}{.}\hspace{.4pt}1109\discretionary{/}{%
}{/}TVCG\hspace{.1pt}\discretionary{.}{%
}{.}\hspace{.4pt}2015\hspace{.1pt}\discretionary{.}{%
}{.}\hspace{.4pt}2467435}}


\end{thebibliography}
\end{document}